\documentclass[journal=jacsat,manuscript=article]{achemso}

\usepackage[version=3]{mhchem} 
\usepackage{csquotes}

\usepackage{physics}

\usepackage{hyperref}
\usepackage[dvipsnames]{xcolor}
\definecolor{lightblue}{RGB}{0,127,255} 

\newcommand{\figref}[2][]{\textcolor{lightblue}{\hyperref[#2]{Figure~\ref*{#2}#1}}}
\newcommand{\eqrefblue}[1]{\textcolor{lightblue}{\hyperref[#1]{Eq.~\ref*{#1}}}}
\newcommand{\methodref}[1]{\textcolor{lightblue}{\hyperref[#1]{Methods}}}

\title{Sensing Electric Currents in an a-IGZO TFT-Based Circuit Using a Quantum Diamond Microscope}

\author{Mayana Yousuf Ali Khan$^\parallel$}
\affiliation{Department of Electrical Engineering and Computer Science, Indian Institute of Science Education and Research, Bhopal}
\author{Pralekh Dubey$^\parallel$}
\affiliation{Department of Physics, Indian Institute of Science Education and Research, Bhopal}
 \author{Lakshmi Madhuri P}
 \affiliation{National Centre for Flexible Electronics, Indian Institute of Technology, Kanpur}
\author{Ashutosh Kumar Tripathi}
\affiliation{National Centre for Flexible Electronics, Indian Institute of Technology, Kanpur}
\author{Phani Kumar Peddibhotla}
\affiliation{Department of Physics, Indian Institute of Science Education and Research, Bhopal}
\email{phani@iiserb.ac.in}
\author{Pydi Ganga Bahubalindruni}
\affiliation{Department of Electrical Engineering and Computer Science, Indian Institute of Science Education and Research, Bhopal}
\email{ganga@iiserb.ac.in}
\date{\today}
   
\begin{document}

\maketitle

\begin{abstract}
The Quantum Diamond Microscope (QDM) is an emerging magnetic imaging tool enabling noninvasive characterization of electronic circuits through spatially mapping current densities. In this work, we demonstrate wafer-level current sensing of a current mirror circuit composed of 16 amorphous-indium-gallium-zinc oxide (a-IGZO) thin-film transistors (TFTs). a-IGZO TFTs are promising for flexible electronics due to their high performance. Using QDM, we obtain two-dimensional (2D) magnetic field images produced by DC currents, from which accurate current density maps are extracted. Notably, QDM measurements agree well with conventional electrical probing measurements, and enable current sensing in internal circuit paths inaccessible via conventional methods. Our results highlight QDM’s capability as a noninvasive diagnostic tool for the characterization of emerging semiconductor technologies, especially oxide-based TFTs. This
approach provides essential insights to fabrication engineers, with potential to improve yield and reliability in flexible electronics manufacturing.
\end{abstract}


\section*{1. INTRODUCTION}
\noindent Magnetic current imaging (MCI) techniques have become indispensable diagnostic tools for integrated circuits (ICs), enabling the detection of faults such as leakage currents, short circuits, and open circuits without physically altering the device.~\cite{knauss2001scanning} These methods involve mapping magnetic fields generated by electric currents flowing through circuit elements, providing critical insights into device functionality, reliability, and failure mechanisms. Traditional MCI techniques employ superconducting quantum interference devices (SQUIDs) and giant magnetoresistance (GMR) sensors. SQUID-based microscopy, while offering excellent magnetic sensitivity, suffers from spatial resolution limitations due to required sensor-to-sample stand-off distances in the order of tens of micrometers.~\cite{fong} Moreover, SQUID systems rely heavily on cryogenic cooling due to their superconducting elements, necessitating complex, expensive, and vibration-sensitive cryo-cooling setups, even in modern cryogen-free implementations.~\cite{dilorio1991manufacturable,horn2019cryogen} In contrast, GMR-based microscopy provides superior spatial resolution, typically reaching sub-micron scales, but has lower magnetic sensitivity in comparison to the SQUID sensors.~\cite{Herrera_2009}
However, both SQUID and GMR microscopes are only sensitive to a single magnetic field component.
\\
\hspace*{10pt} To overcome these limitations, the Quantum Diamond Microscope (QDM) has emerged as an advanced diagnostic technique, offering high-resolution, wide-area, non-invasive magnetic imaging at room temperature.~\cite{levine,glenn,Scholten_2021} Recent advancements in NV center control and sensor-sample proximity have further enhanced the spatial resolution and magnetic sensitivity of QDM systems.~\cite{hudak2023atomically,xu2025minimizing} QDM utilizes nitrogen-vacancy (NV) colour centres embedded within a diamond substrate to detect and map magnetic fields generated by electrical currents. Owing to the projective magnetic microscopy (PMM) as well as vector magnetic field microscopy (VMM) capabilities, \cite{glenn} QDM has successfully been applied to various semiconductor technologies. It has been implemented to sense currents or detect defects in 555-timer ICs based on bipolar junction transistors (BJTs),\cite{Kehayias:2022} a silicon device doped with phosphorus,~\cite{Basso:2023}  silicon photovoltaic (PV) devices,~\cite{scholten2022imaging} monolayer graphene structures, carbon nanotubes \cite{Tetienne:2017,chang,Zhong_2024} and three-dimensional SiGe multilayer structures.~\cite{Garsi:2024}  Furthermore, magnetic field and current density maps have been extracted from ring oscillator (RO) clusters in field-programmable gate arrays (FPGAs) following decapsulation,\cite{Turner:2020} and three-dimensional current distributions through metal interconnects, vias, and power rails have been imaged in a graphics processing unit (GPU).\cite{Oliver} These examples underscore the broad applicability of QDM as a noninvasive diagnostic platform for diverse material systems and device architectures.
\begin{figure*}[!h]
\centering
\includegraphics[width=1.0\textwidth]{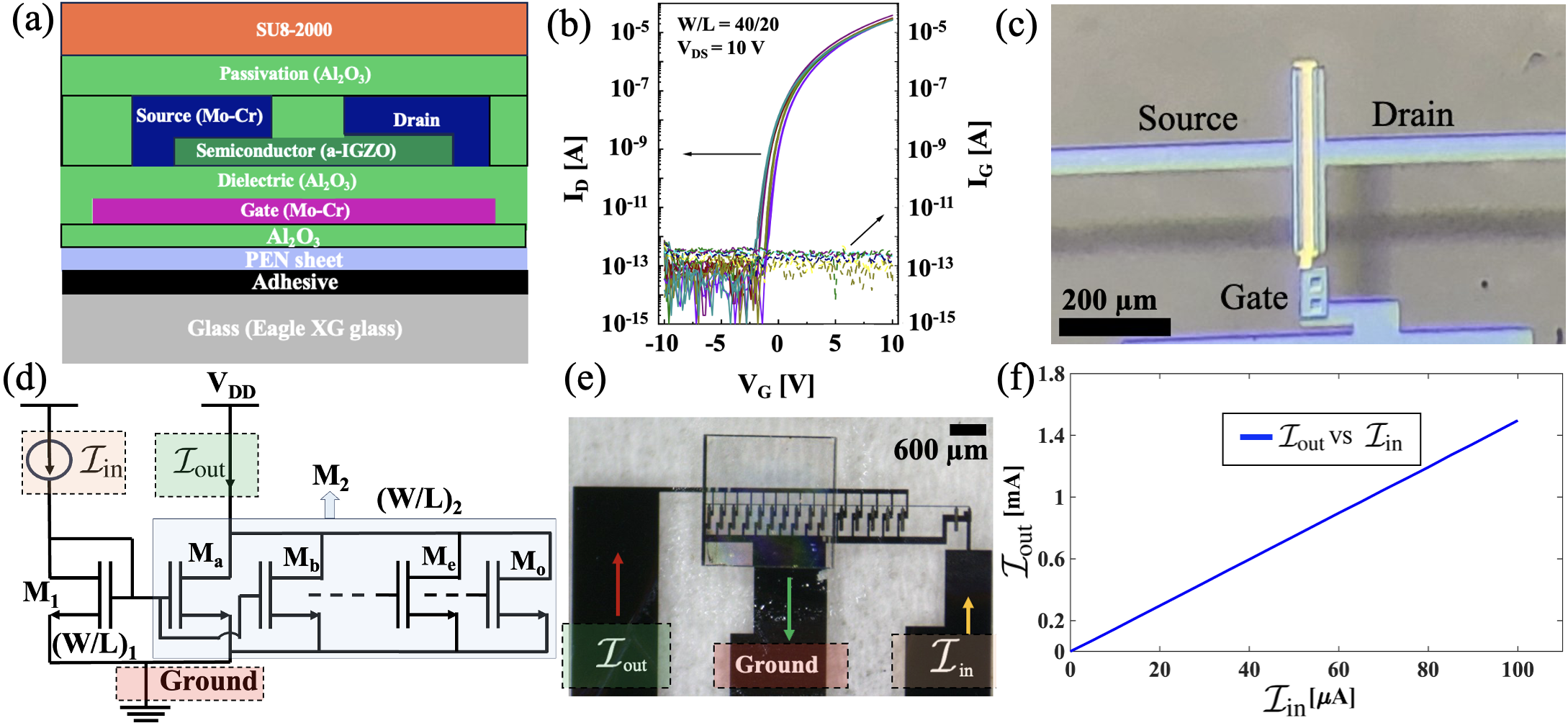}
\caption{(a) Cross-sectional schematic of an a-IGZO TFT. (b) Transfer characteristics for 6 TFTs over a 4” x 4” substrate. The dimensions of the transistor channel are width ($W$) = 40 $\mu$m and length ($L$) = 20 $\mu$m. (c) A zoomed-in micrograph of an individual a-IGZO transistor with channel dimensions $W$ = 320 $\mu$m and $L$ = 20 $\mu$m. (d) Schematic of the current mirror circuit. (e) Micrograph of the current mirror circuit with one transistor at the input, and 15 ($M_a$, $M_b$, - - - - -, $M_o$) transistors at the output. (f) Measured response of the current mirror circuit, 
showing a mirroring ratio of 14.92.}
\label{fig:fig1circuit}
\end{figure*} 
\\
\hspace*{10pt} Despite significant progress, QDM-based current imaging studies have mainly focused on silicon-based devices or circuits. There is a notable absence of similar diagnostic studies for circuits or systems with flexible electronics, which find potential applications in various domains, including displays, RFID, NFC tags, biomedical wearable devices, eskin, energy harvesting devices, image recognition, and gas detection.\cite{nakano,zhang2022ultraviolet,kim,wieczorek2023thin,meister202049,monga2024flexible,geng2023thin,zhao,shri,velazquez2024tunable,jin2015water,jang2022amorphous,kim2021stress,jia,qi2025integrated,sen2022probing,kim2018toward,chu2022power} Among the semiconductor technologies that facilitate flexible electronics, oxide TFTs offer several advantages. These advantages include higher electron mobility ($>$ 10 cm$^2$/Vs), large-area uniformity, low off-state currents, sharp sub-threshold swings, superior electrical stability,~\cite{nomura2004room,reuss2005macroelectronics} and enhanced reliability and mechanical durability.~\cite{han2021mechanical,shi2024approach} These characteristics make oxide (a-IGZO) TFT technology an optimal choice for developing large-area, lightweight, bendable, and stretchable electronic devices. It is worth mentioning that the oxide (a-IGZO) TFT technology is relatively in the emerging state compared to the standard silicon-based semiconductor technology.~\cite{troughton2019amorphous}  Issues such as yield reduction due to material processing inconsistencies and layer misalignment, particularly gate-to-source/drain, often introduce additional parasitic capacitance and degrade device performance.\cite{bahubalindruni2013transparent,Luo_2023} As such, accurate diagnostic tools capable of identifying and characterizing current paths and detecting potential faults in a noninvasive way with high spatial resolution become critical. QDM, which offers high magnetic sensitivity and spatial resolution, is uniquely positioned to address these diagnostic challenges in oxide TFT technologies.
\\
\hspace*{10pt} This manuscript presents non-invasive current sensing using QDM, in a two-transistor current mirror (CM) circuit implemented with a-IGZO TFTs. After applying a suitable stimulus to the CM, magnetic field maps are obtained at two different
conducting paths in the circuit in order to validate the suitability of QDM for noninvasive current sensing. This includes the paths that are not directly accessible for measurement with other conventional circuit testing methods. Therefore, this work can provide crucial information to the fabrication engineers, in order to improve process yield in fabrication.

\section*{2. RESULTS AND DISCUSSION}
\textbf{2.1. Oxide TFT Technology:} Cross-sectional schematic of an individual oxide(a-IGZO) TFT  is shown in \figref{fig:fig1circuit}a. 
IGZO TFTs and circuits were fabricated using low temperature processes, not exceeding  $150 ^{\circ}$C, making them compatible with glass as well as polymer substrates. A 125 $\mu$m thick heat stabilized PEN sheet was bonded on an Eagle XG glass substrate of 0.7 mm thickness using a heat sensitive adhesive. Further, a buffer layer of 100 nm thick aluminium oxide $(Al_2O_3)$ layer was deposited on bonded PEN substrate by atomic layer deposition (ALD) at $150 ^{\circ}$C. For gate electrode, 100 nm Mo-Cr (10\% Cr) was sputter deposited. Gate insulator consists of 80 nm thick $Al_2O_3$ film deposited by ALD at $150 ^{\circ}$C. Semiconductor layer was made of a $\sim 25$ nm thin film of Indium Gallium Zinc Oxide (IGZO) deposited using sputtering from an IGZO target with a molecular ratio of 1:1:1 under Ar gas at a process pressure of 1 mTorr. All the above layers were patterned by wet-etching and standard photolithography. Source and drain electrodes were defined by lift-off of a 100 nm thick film of Mo-Cr. An 80 nm thick layer of $Al_2O_3$ was deposited using ALD at $150 ^{\circ}$C to passivate devices and circuits. Measurement pads were opened using standard wet-etching process. In order to provide mechanical robustness as well as additional device stability, a 1.2 $\mu$m thick layer of SU8-2000 was coated and patterned to open measurement pads. Finally, a baking step at $150 ^{\circ}$C for two hours under clean dry air was executed. 
\\
\hspace*{10pt} \figref{fig:fig1circuit}b shows transfer characteristics for 6 TFTs fabricated on a $4'' \times 4''$ substrate.  Their electrical parameters  are: mobility ($\mu_e$) of $8.8\,\pm\,0.05$ cm$^2$/Vs, a turn-on voltage of $-1.6\,\pm\,0.4$ V, and a subthreshold swing of $300\,\pm\,40 $ mV/dec.  Gate leakage current below 1 pA and an ON/OFF ratio $>$ $10^8$ was achieved. Despite the constraints associated with low-temperature processing for flexible substrates, the achieved performance is adequate to realize  functional circuits. 
\\
\hspace*{10pt} \textbf{2.2. Circuit Design.}
The CM is an essential building block in ICs, which finds applications in  biasing, current amplification, and serving as an active load. \figref{fig:fig1circuit}d illustrates a schematic of the CM circuit with transistor channel dimensions of width ($W$)  and length ($L$)  in $\mu$m. The aspect ratios $\bigl[W/L]$ of the input transistor ($M_1$) and output ($M_2$) transistors are $\bigl[320/20\bigr]$ and $\bigl[4800/20\bigr]$, respectively. \figref{fig:fig1circuit}e shows a micrograph of the current mirror circuit with a unit TFT channel dimensions of width $W$ = 320 $\mu$m and  length $L$ = 20 $\mu$m. \figref{fig:fig1circuit}c shows the micrograph of the unit transistor. Here, the output transistor is implemented as a parallel combination of 15 unit transistors ($M_a$, $M_b$, - - - - -, $M_o$), while the input transistor is a single unit transistor. The relation between $\mathcal{I}_{\text{in}}$ and $\mathcal{I}_{\text{out}}$ given by :
\begin{equation}\label{eq1}
    \mathcal{I}_{\text{out}} = \mathcal{I}_{\text{in}} \frac{(W/L)_{2}}{(W/L)_{1}}.
\end{equation}
Here, $(W/L)_{1}$ and $(W/L)_{2}$ are the aspect ratios of the transistors $M_{1}$ and $M_{2}$, respectively. Maintaining the same channel length ($L$) across all transistors in the circuit makes $\mathcal{I}_{\text{out}}$ depend only on the  width ($W$) of both the transistors. Since, there are 15 unit transistors connected in parallel (\figref{fig:fig1circuit}e), the width of $M_2$ is 15 times that of $M_1$ resulting a mirror ratio of 15. Under ambient conditions, this circuit is experimentally characterized by sweeping the current $\mathcal{I}_{\text{in}}$ from 0.1 $\mu$A to 100 $\mu$A. \figref{fig:fig1circuit}f demonstrates a linear relationship  between $\mathcal{I}_{\text{in}}$ and $\mathcal{I}_{\text{out}}$, with a mirroring ratio of 14.92 at $V_{\text{DD}}$= 2 V.
The slight deviation from the ideal ratio of 15 can be attributed to several non-idealities in the fabrication and device characteristics, such as mismatch, process non-uniformity, and parasitic effects from interconnects.\cite{bahubalindruni2013transparent} 
\begin{figure*}[!h]
\centering
\includegraphics[width=1\textwidth]{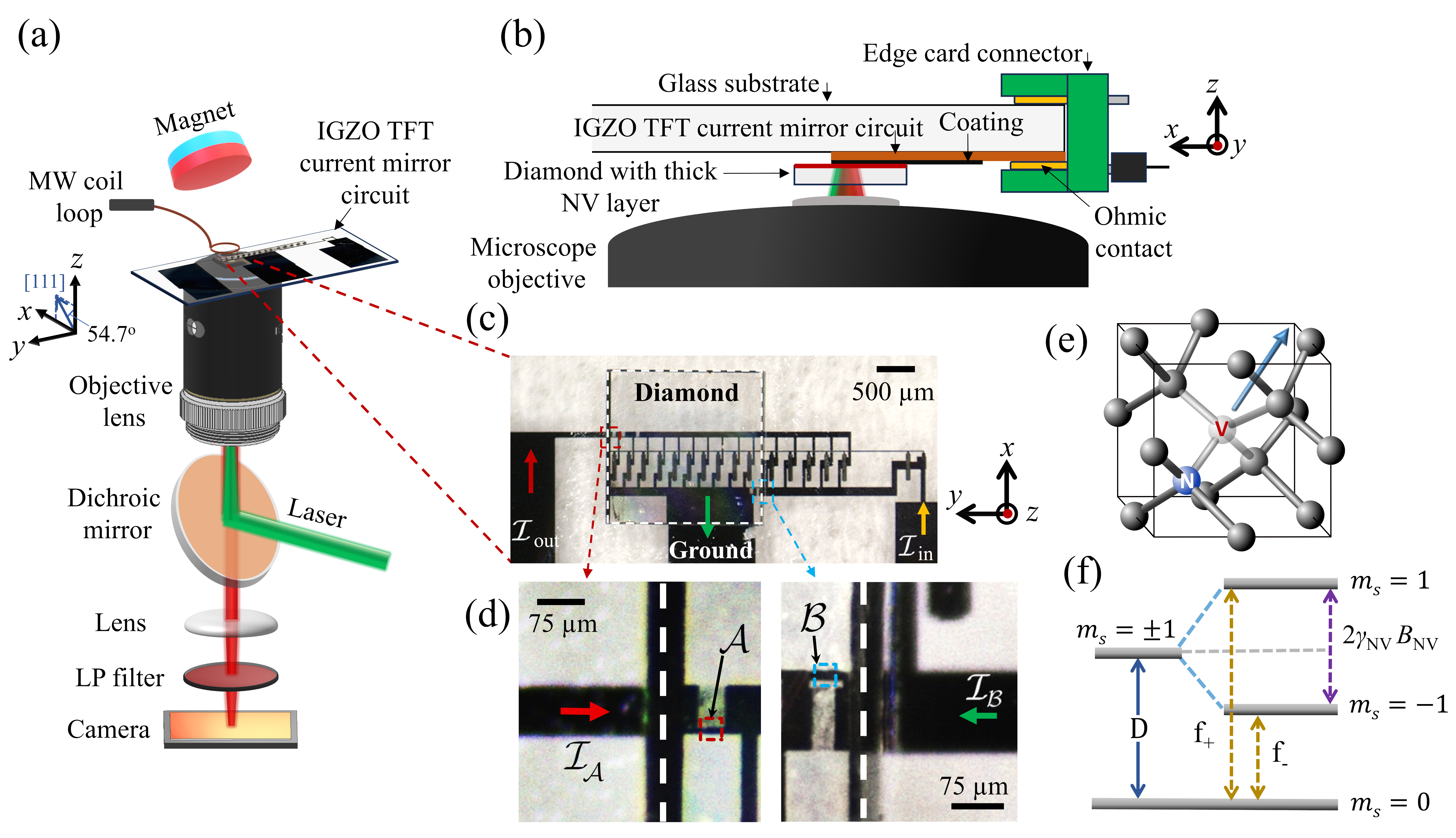}
\caption{(a) Schematic diagram of the QDM setup used for sensing DC electric currents in the a-IGZO TFT-based CM circuit. (b) The diagram presents a cross-sectional side view of the setup. The diamond is placed on top of the DUT (CM circuit) such that the surface containing the NV layer is in contact with the circuit. Laser beam, delivered through the objective lens, illuminates the NV layer at the region of interest. The card edge connector facilitates current stimulus to the circuit by establishing a reliable connection to the conducting pads of the circuit. (c) The micrograph shows the diamond placed on top of the CM circuit, which consists of 16 TFTs. The current stimulus is applied through the $\mathcal{I}_{\text{in}}$ conducting pad, and the application of the voltage $V_{\text{DD}}$ to the $\mathcal{I}_{\text{out}}$ pad scales the output current (see \figref{fig:fig1circuit}d). (d)  A zoomed-in micrograph of the circuit displaying a close-up view of the region $\mathcal{A}$ near the $\mathcal{I}_{\text{out}}$ terminal, and a close-up view of the region $\mathcal{B}$ near the ground terminal. We obtained magnetic field images of the current paths in these two regions. The width of the current paths in the regions $\mathcal{A}$ and $\mathcal{B}$ are $\sim$9.5 $\mu$m and $\sim$15 $\mu$m, respectively. (e) Defect structure of an NV center in diamond. The gray spheres represent the carbon atoms, while the blue and white spheres together represents the NV center. An external magnetic field $\vb*{B}$ acting along the NV axis is also shown with a blue arrow. (f) The energy level diagram of the ground-state spin of the NV center. Here, $D$ is the zero-field splitting (ZFS) between $\ket{m_s=0}$ and $\ket{m_s=\pm 1}$ spin states.  With the application of a magnetic field, the degeneracy between $\ket{m_s=\pm 1}$ is lifted due to the Zeeman effect, resulting in spin transition frequencies $f_-$ for the transition between  $\ket{m_s=0}$ and $\ket{m_s= - 1}$, and $f_+$ for the transition between  $\ket{m_s=0}$ and $\ket{m_s= - 1}$ spin states.}
\label{fig:schematic} 
\end{figure*} 
\\
\hspace*{10pt} \textbf{2.3. Experimental Setup.} 
\figref{fig:schematic}a shows a schematic of the experimental setup, highlighting its operational aspects (see also the ``\methodref{sec:methods}'' section). The sensor chip is a 2 mm$\times$2 mm$\times$0.5 mm single-crystal CVD diamond substrate, with $[110]$ edge orientations and a $\{100\}$ front facet. The NV defects are embedded within $\sim$2.5 $\mu$m-thick isotopically purified $[^{12}\mathrm{C}]$ layer, with a NV concentration of approximately $\sim$1 ppm. The diamond is placed directly on top of the DUT (CM circuit) with the NV layer in contact with the current-carrying circuit (\figref{fig:schematic}b). A 532 nm laser beam is focused onto the NV layer using an air objective (40$\cross$ 0.55 NA) leading to a spot size of 30 $\mu$m diameter at the region of interest. The excited NV defects emit red fluorescence (wavelength 637-800 nm), which is collected through the same objective lens. To drive the NV spin transitions between the spin states, microwaves (MW) are delivered via a loop antenna placed in close proximity to the diamond, thereby ensuring efficient delivery of the microwaves to the NV centers (\figref{fig:schematic}a). Finally, the fluorescence collected is filtered with a 650 nm longpass filter, and is imaged onto a sCMOS camera. The fluorescence images captured by the camera are then processed to construct the magnetic field map resulting from the current flowing through the conducting path in the region of interest in the circuit. 
\\
\hspace*{10pt} NV centers, employed in QDM, are spin-1 quantum defects in diamond exhibiting optically addressable ground-state spin which is sensitive to magnetic field. External magnetic field induces Zeeman shift of the NV spin states, detectable via optically detected magnetic resonance (ODMR) spectroscopy.~\cite{Gruber:1997,Doherty:2013} 
By analyzing Zeeman shifts across the NV layer, spatially resolved two-dimensional (2D) magnetic field maps are produced, offering detailed magnetic field images of the device. 
\\ 
\hspace*{10pt} We adopt PMM ODMR technique,\cite{glenn} for mapping the Oersted fields generated by the CM circuit (see also the ``\methodref{sec:methods}'' section). A static bias magnetic field of $B_{\text{bias}}\approx$ 2 mT, generated by a NdFeB permanent magnet, is applied along the [111] crystallographic direction, which corresponds to one of the four possible orientations of the NV symmetry axes in diamond  (\figref{fig:schematic}e). 
The ground-state spin Hamiltonian, for this NV orientation class, representing  the Zeeman interaction with the bias field and the Oersted field is:
\begin{equation}
    \label{Eq:Hamiltonian1}
    \mathcal{\hat{H}}=D \hat{S}_z^2+\gamma_{{\text{NV}}}(B_{\text{bias}}+{B_{\text{sample}}})\hat{S}_z,
\end{equation}
where $D=2870$ MHz is the zero-field splitting (ZFS), $\gamma_{{\text{NV}}}=$ 28 MHz/mT is the gyromagnetic ratio of NV spin, $B_{\text{bias}}$ is the magnitude of the external bias magnetic field along the NV axis and $B_{\text{sample}}$ is the projection of the magnetic field along the NV axis due to the current flowing in the CM circuit.
Thus, the effective field along the NV axis, i.e.,  $B_{\text{NV}} = B_{\text{bias}} + B_{\text{sample}}$, causes the Zeeman splitting between the $m_s = \pm1$ states. It shifts the resonance frequencies between $\ket{m_s = 0}$ and $\ket{m_s = 1}$ ($f_+$) and between $\ket{m_s = 0}$ and $\ket{m_s = -1}$ ($f_-$) by $\pm \gamma B_{\text{NV}}$ relative to ZFS (\figref{fig:schematic}f). ODMR spectra are obtained by sweeping the MW frequency across both the NV spin resonances in discrete frequency steps and capturing the fluorescence images at every frequency step using a camera. For mapping the magnetic field, ODMR spectrum is analyzed for every pixel of the recorded fluorescence images and fitted with a Lorentzian function to determine the frequencies ($f_-$ and $f_+$) corresponding to the ODMR dips. Subsequently, the effective field $B_{\text{NV}}$ is calculated using $(f_{+}-f_{-})$/(2$\gamma_{\text{NV}}$). The magnetic field $B_{\text{bias}}$ measured under open-circuit condition is then subtracted from the extracted $B_{\text{NV}}$ to generate the 2D magnetic field map corresponding to the current flowing through the circuit (\figref{fig:mf_image}).
\\
\begin{figure*}[h!]
\centering
\includegraphics[width=1\textwidth]{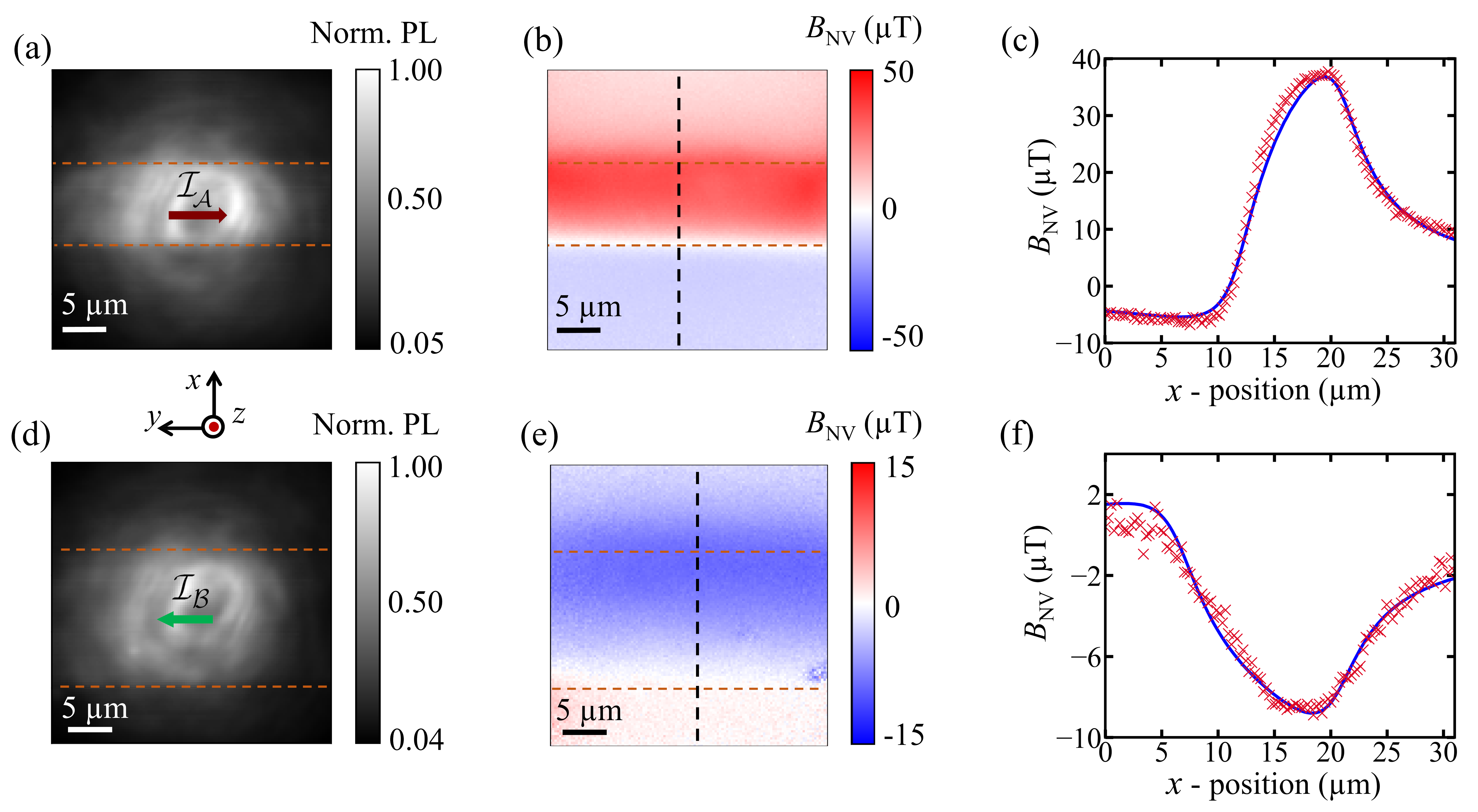}
\caption{Imaging of the Oersted magnetic field produced by the current-carrying conducting strips. (a) The photoluminescence (PL) image of the region $\mathcal{A}$.  (b) Image of the magnetic field ($B_{\text{NV}}$) due to the current $\mathcal{I_A}$ flowing through the conducting path in the region $\mathcal{A}$. (c) The magnetic field profile is measured along the black dashed line in (b). Red crosses represent experimental data, and the blue curve is a fit (see ``\methodref{sec:methods}'') providing a standoff distance of 2.3 $\mu$m. (d) The PL image of the region $\mathcal{B}$. (e) Image of the  magnetic field ($B_{\text{NV}}$) due to the current $\mathcal{I_B}$ flowing through the conducting path in the region $\mathcal{B}$. (f) The magnetic field profile is measured along the black dashed line in (e). The blue curve is a fit to the experimental data (red crosses) giving a stand-off distance of 2.8 $\mu$m. For the images shown in (a), (b), (d) and (e), the edges of the conducting strips are marked by orange dashed lines.}
\label{fig:mf_image} 
\end{figure*} 
\hspace*{10pt} \textbf{2.4. Measurement of Current-Induced Magnetic Field.} Magnetic field measurements are performed at two separate regions of the CM circuit to verify the proper mirroring of the current: (i) the conductive path in the region $\mathcal{A}$ near the $\mathcal{I}_{\text{out}}$ terminal, and (ii) the conductive path in the region $\mathcal{B}$ near the ground terminal. These regions are shown and labeled in \figref{fig:schematic}d. Near the region $\mathcal{A}$, there are 15 transistors ($M_a$, $M_b$, - - - - -, $M_o$) connected in parallel, each drawing $\mathcal{I}_{\text{in}}$. This scales the output current by a factor of 15, i.e., the maximum possible $\mathcal{I}_{\text{out}}$ is expected to flow through this conducting path. 
Thus, the magnetic field measurement in this region will verify the proper mirroring of the current by the circuit. We note that the direction of the current flow is along $-y$ direction of the coordinate system. 
\figref{fig:mf_image}a and \figref{fig:mf_image}d show the photoluminescence (PL) images captured by the camera at regions $\mathcal{A}$ and $\mathcal{B}$, respectively. The corresponding $B_{\text{NV}}$ magnetic field map extracted from region $\mathcal{A}$ is shown in \figref{fig:mf_image}b. The current carrying strip has a width of $w_{\mathcal{A}}\sim$9.5 $\mu$m, and the strip is marked by orange dashed lines in the 2D plots. Linecut indicated by the black dashed line in \figref{fig:mf_image}b is shown in \figref{fig:mf_image}c to illustrate the profile of the measured field. The other region of investigation, $\mathcal{B}$, near the ground terminal, is located after six TFTs (including the single unit input TFT at $\mathcal{I}_{\text{in}}$ terminal) from the input terminal. The current through the conducting path in this region is, thus, expected to scale by a factor of 6 due to the presence of the six TFTs, each outputting $\mathcal{I}_{\text{in}}$ current. This part of the circuit is inaccessible for analysis by the conventional electrical probing technique, i.e., using a probe station, because the passivation layer prevents contact with the conducting strip. On the other hand, the Oersted fields due to the current flowing through the conducting path remains accessible for characterization by the QDM setup. \figref{fig:mf_image}e shows the extracted $B_{\text{NV}}$ field map of the region $\mathcal{B}$. Note that the direction of the current flow is along $y$ direction of the coordination system. The width of the current carrying strip, marked by orange dashed lines, is $w_{\mathcal{B}}\sim$15.5 $\mu$m (\figref{fig:mf_image}a). \figref{fig:mf_image}f plots a line cut of the magnetic field $B_{\text{NV}}$ for the black dashed line in \figref{fig:mf_image}e. 
\\
\hspace*{10pt} Since the current path is parallel to the $y$-axis (see \figref{fig:mf_image}c) in both the regions ${\mathcal{A}}$ and ${\mathcal{B}}$, we focus on the non-zero components $B_x$ and $B_z$ of the Oersted magnetic field. However, our PMM approach measures only the projection of the field along the NV axis, $B_{\text{NV}}$, which can be expressed in terms of $B_x$ and $B_z$ as:
\begin{equation}
    \label{Eq:Bnv}
B_{\text{NV}}=B_x\sin{\theta}\cos{\phi} + B_z\cos{\theta},
\end{equation}
where $\phi$  and $\theta$ are the azimuthal and polar angles defining the NV axis direction (\figref{fig:schematic}a).
To estimate the stand-off distance ($h$) between the NV layer and the current carrying strip, we fit the measured magnetic field to the theoretically predicted current-induced magnetic field by the Biot-Savart law, using $h$ as the free parameter. The line cut plots are fitted using Eq. \ref{eq:BNV_prime}, which also takes the NV layer thickness ($\sim$2.5 $\mu$m) into consideration (see ``\methodref{sec:methods}''), and the fitted plots are shown as blue curves in \figref{fig:mf_image}c and \ref{fig:mf_image}f. Thus, the stand-off distances are estimated to be $h_{\mathcal{A}} = 2.3$\space$\mu$m and $h_{\mathcal{B}} = 2.8$\space$\mu$m. These values are used in reconstructing the $K_x$ and $K_y$ current density images at the regions we investigate in this study.
\begin{figure*}[h]
\centering
\includegraphics[width=1\textwidth]{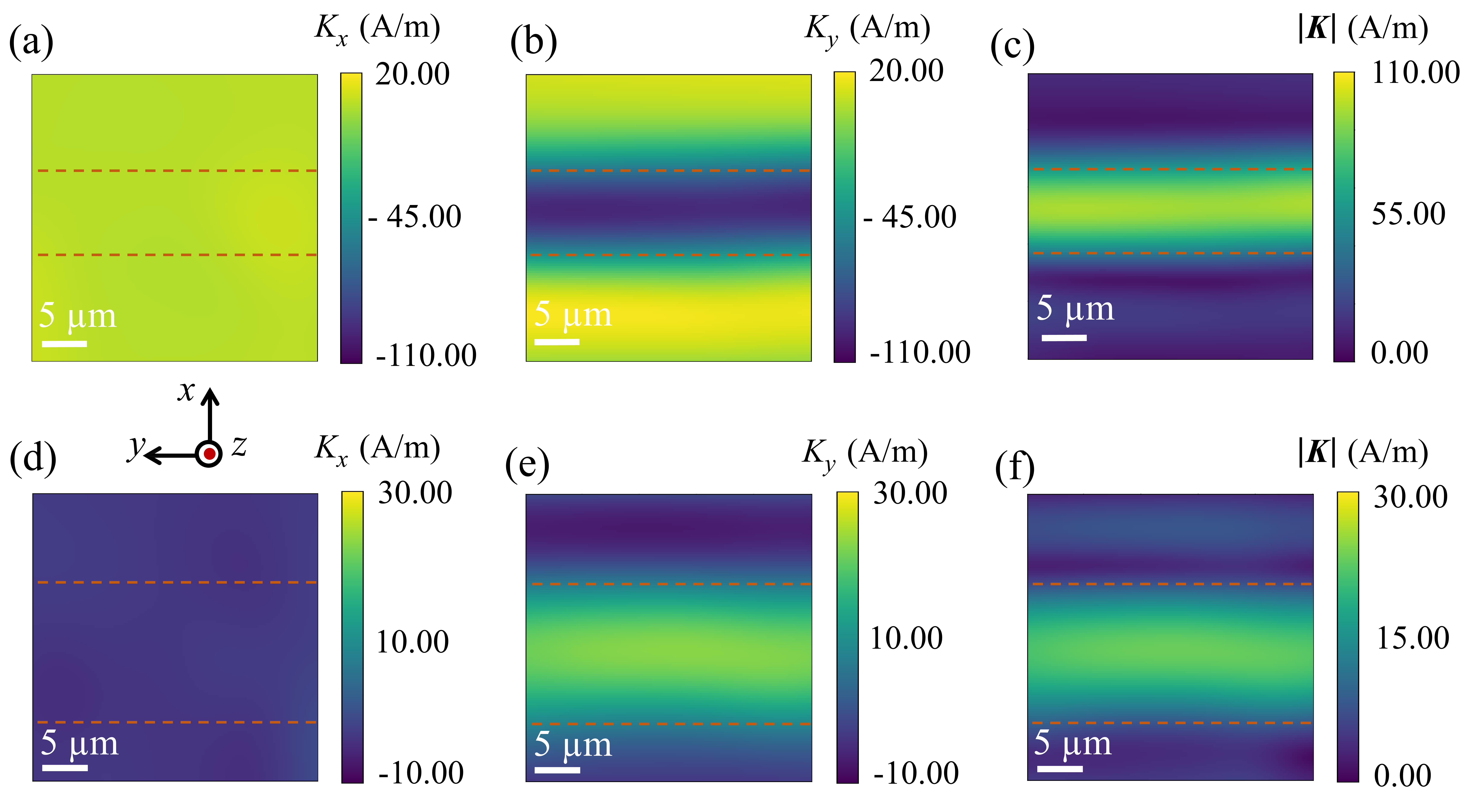}
\caption{Reconstructed current density images.
(a, d) Images of the $K_x$ component of the current density of the regions $\mathcal{A}$ and $\mathcal{B}$. (b, e) Images of the $K_y$ component of the current density. (c, f) Images of the resultant current density magnitude $\abs{\vb*{K}}$. The edges of the conducting strips are marked by orange dashed lines in all the images.}
\label{fig:current_density_vdd} 
\end{figure*}
\hspace*{10pt} \textbf{2.5. Reconstruction of Current Density.}
The reconstruction of the current density is achieved following the method described by Roth et al. \cite{roth}. We perform inversion of Biot-Savart's law in Fourier space, thus enabling us to infer the current density $\vb*{K}(x,y)$ from the magnetic field. The reconstruction process is discussed in detail in the Supporting Information.
\figref{fig:current_density_vdd} shows the vector components $K_x$ and $K_y$ of the reconstructed current density as well as its magnitude $\abs{\vb*{K}}=\sqrt{K_x^2+K_y^2}$ for the regions $\mathcal{A}$ and $\mathcal{B}$. The negative value of $K_y$ at region $\mathcal{A}$ indicates that the current flows in the $-y$ direction, whereas the positive value of $K_y$ at region $\mathcal{B}$ suggests that the current flows along the $+y$ direction. This is consistent with the actual direction of the current flow in both the regions of the current mirror circuit. 
\\ 
\hspace*{10pt} To validate the accuracy of the reconstruction, the current density $\abs{\vb*{K}}$ in the 2D plots, shown in \figref{fig:current_density_vdd}c and \ref{fig:current_density_vdd}f, is integrated over the width of the conducting strip to obtain the magnitude of the current flowing through the two regions. For this analysis, 1D cut-line plots are obtained to find the mean current density values at $\mathcal{A}$ and $\mathcal{B}$ 
(detailed in Supplementary Information section VII). The mean current density value at region $\mathcal{A}$ is found to be 79.5 A/m $\pm$ 0.5 A/m, while at region $\mathcal{B}$ the mean current density value is found to be 17.3 A/m $\pm$ 0.2 A/m.
Using these mean values, the magnitude of the currents flowing through the two conductive paths ($\mathcal{A}$ and $\mathcal{B}$) in the CM circuit is obtained and compared with the values obtained via (i) theoretical analysis, (ii) circuit simulation using EDA tools (Cadence Virtuoso), and (iii) conventional testing (see ``\methodref{sec:methods}'') as summarized in Table \ref{tableCMP}. 
For the QDM method, the magnitude of the extracted current is 755.2 $\mu$A in the region $\mathcal{A}$ and 268.1 $\mu$A in the region $\mathcal{B}$. This translates to a percentage error of 1.2$\%$ relative to conventional measurements in the region $\mathcal{A}$, which improves to 0.68$\%$ when compared with simulation results. In contrast, the current in  region $\mathcal{B}$ cannot be measured using the conventional technique due to the  insulating passivation layer and limited physical accessibility for electrical probe. As a result, QDM measurements in region $\mathcal{B}$ are validated against simulation result, yielding a percentage error of 10.6$\%$. The comparatively larger error in region $\mathcal{B}$ may stem from the theoretical current assumed for error estimation being higher than the actual current flowing through the region.
\begin{table*}[h!]
\centering
\caption{Comparative analysis of four distinct testing methodologies to evaluate the performance of the CM circuit}
\label{tableCMP}
\setlength{\tabcolsep}{3.5pt} 
\begin{tabular}{|p{0.8cm}|p{1.8cm}|p{2.6cm}|p{2.6cm}|p{2.6cm}|p{2.6cm}|c|c|} 
\hline
S. No. & \hspace*{2pt}Input current $\mathcal{I}_{\text{in}}$ & Current \newline (Expected) & Current \newline(simulation) & Current \newline(conventional) & Current (QDM) & \multicolumn{2}{c|}{\% Error} \\ 
\cline{7-8}
 & & & & & & \small $*$ & \small $\dag$ \\ 
\hline
\hspace*{2pt}1 & \hspace*{12pt}50 $\mu$A & $I_{\mathcal{A}}$ = 750 $\mu$A & $I_{\mathcal{A}}$ = 750 $\mu$A & $I_{\mathcal{A}}$ = 746.1 $\mu$A & $I_{\mathcal{A}}$ = 755.2 $\mu$A & 1.2 & 0.68 \\ 
\hline
\hspace*{2pt}2 & \hspace*{12pt}50 $\mu$A & $I_{\mathcal{B}}$ = 300 $\mu$A & $I_{\mathcal{B}}$ = 300 $\mu$A & -- & $I_{\mathcal{B}}$ = 268.2 $\mu$A & -- & 10.6 \\ 
\hline
\end{tabular}

\vspace{2mm}
\parbox{\linewidth}{
\footnotesize
\textit{(\enquote{$*$} denotes the percentage error obtained while comparing the current measured using the QDM to that measured using the conventional method \\
\enquote{$\dag$} denotes the percentage error obtained while comparing the current measured using the QDM to the theoretically expected current)}
}
\end{table*}
\begin{figure*}[h]
\centering
\includegraphics[width=1\textwidth]{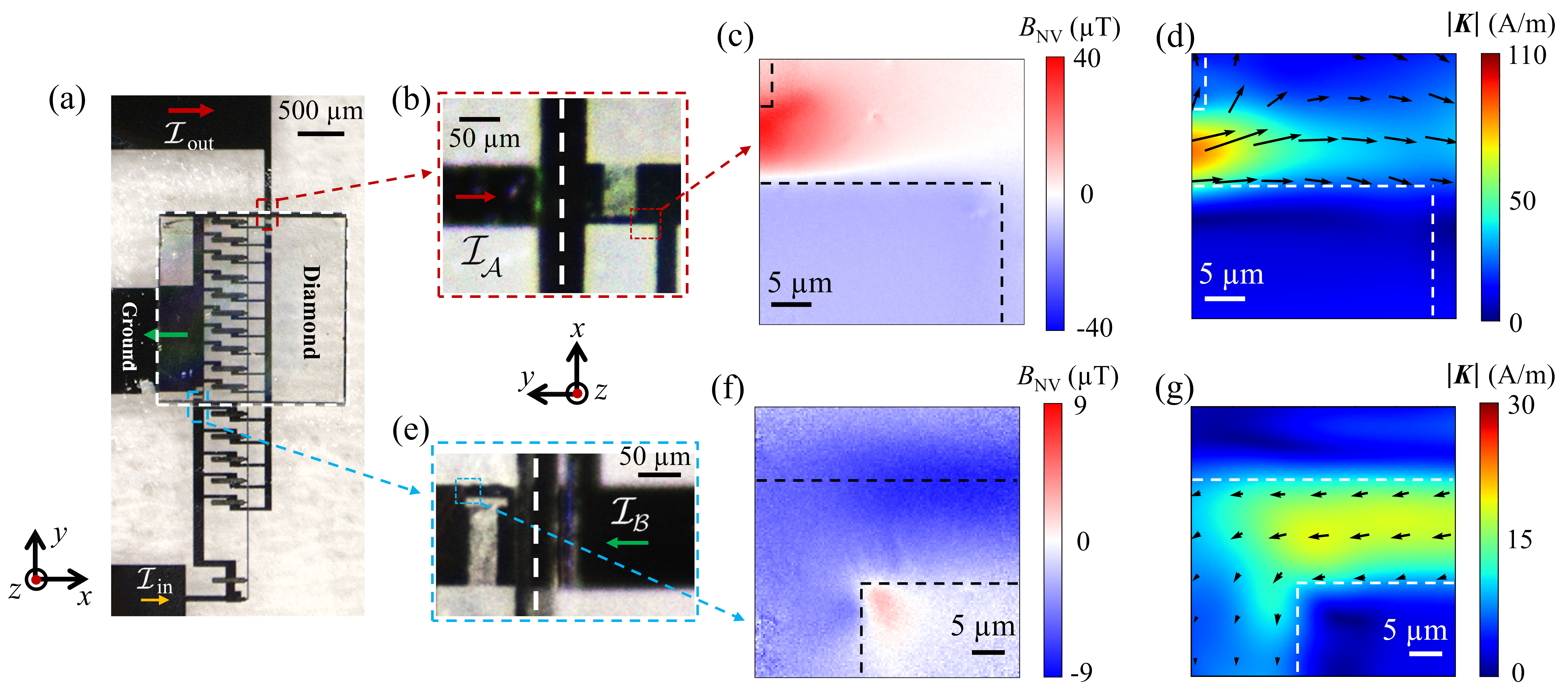}
\caption{Reconstructed current density images near the corner regions of the circuit. (a) The micrograph of the a-IGZO TFT CM circuit. (b, e) The zoomed-in images at the regions $\mathcal{A}$ and $\mathcal{B}$ focusing on the corner regions chosen for this study. The red and green solid arrows show the direction of the current flow. (c, f) Magnetic field maps obtained at the respective corner regions. Black dashed lines outline the conductor edges (d, g) Reconstructed current density $\abs{\vb*{K}}$ maps at the corner regions near $\mathcal{A}$ and $\mathcal{B}$, obtained from the respective magnetic field maps. White dashed lines outline the conductor edges. The black arrows represent the current density vector $\vb*{J}$ indicating the direction of the current flow. The length of each arrow is proportional to $\abs{\vb*{K}}$. For plotting the arrows, a threshold of $\abs{\vb*{K}} > 12\,\mathrm{A/m}$ in (d) and $5\,\mathrm{A/m}$ in (g) is set.}
\label{fig:uorl} 
\end{figure*}
\\
\hspace*{10pt}To further advance our investigation and as a key step towards application to more complex a-IGZO TFT-based circuit geometries, we map the current densities at the corner regions of the conducting paths near the regions $\mathcal{A}$ and $\mathcal{B}$ (\figref{fig:uorl}b and \ref{fig:uorl}e). Here, we focus on the variation in the current density as the direction of the current flow changes through these regions. \figref{fig:uorl}c and \ref{fig:uorl}f illustrate the corresponding $B_{\text{NV}}$ magnetic field plots acquired at these locations. Current density images are reconstructed using the approach previously described in \figref{fig:current_density_vdd}. Additionally, the black arrows representing the current flow paths are derived directly from the vector components $K_x$ and $K_y$. \figref{fig:uorl}d and \ref{fig:uorl}g clearly show a sudden change in $\abs{\vb*{K}}$ at the corner regions, which is also evident from the distribution of the magnetic field in those regions. The changes in electric current flow arise not only due to a change in the direction of current but also because of varying path widths. Specifically, as the electric current transitions from a narrower conducting path to a wider region, a noticeable reduction in $\abs{\vb*{K}}$ is observed (\figref{fig:uorl}d and \ref{fig:uorl}g), which is represented by the reduced size of the arrows. Additionally, in the plots, one can observe the apparent presence of current density outside the conducting strip. This artifact is attributed to optical aberrations, \cite{scholten2022imaging,Tetienne:2019} and inherent noise from the single-axis FFT-based current density reconstruction process. \cite{Broadway:2020} These measurements would help in visualizing the current distribution as it flows through the different paths in the ciruit. It could be useful in resolving intricate current paths in geometrically complex current routing in a-IGZO TFT based circuits.
\section*{3. CONCLUSION}
This work demonstrates the efficacy of wide-field diamond magnetometry in performing non-invasive imaging of current flow in an a-IGZO TFT based CM circuit.
The QDM is validated at two distinct paths  ($\mathcal{A}$ and $\mathcal{B}$) within the circuit, including an internal conducting path (region $\mathcal{B}$) that is inaccessible through standard electrical probing. The magnitude of current extracted from the reconstructed current density maps matches closely with theoretically expected values, circuit simulations and conventional electrical measurements (only $\mathcal{A}$), with an error margin within 1–10\%.  Additionally, the current density plots at the corner regions of the conducting paths can also be extracted from QDM. Overall, this study establishes QDM as a novel and powerful diagnostic tool for evaluating oxide semiconductor-based circuits. It not only complements  conventional electrical measurements but also opens new pathways for wafer-scale inspection, failure analysis, and yield optimization in emerging flexible and transparent electronics. 
\section*{METHODS}
\label{sec:methods}
\begin{figure}
    \centering
    \includegraphics[width=1\linewidth]{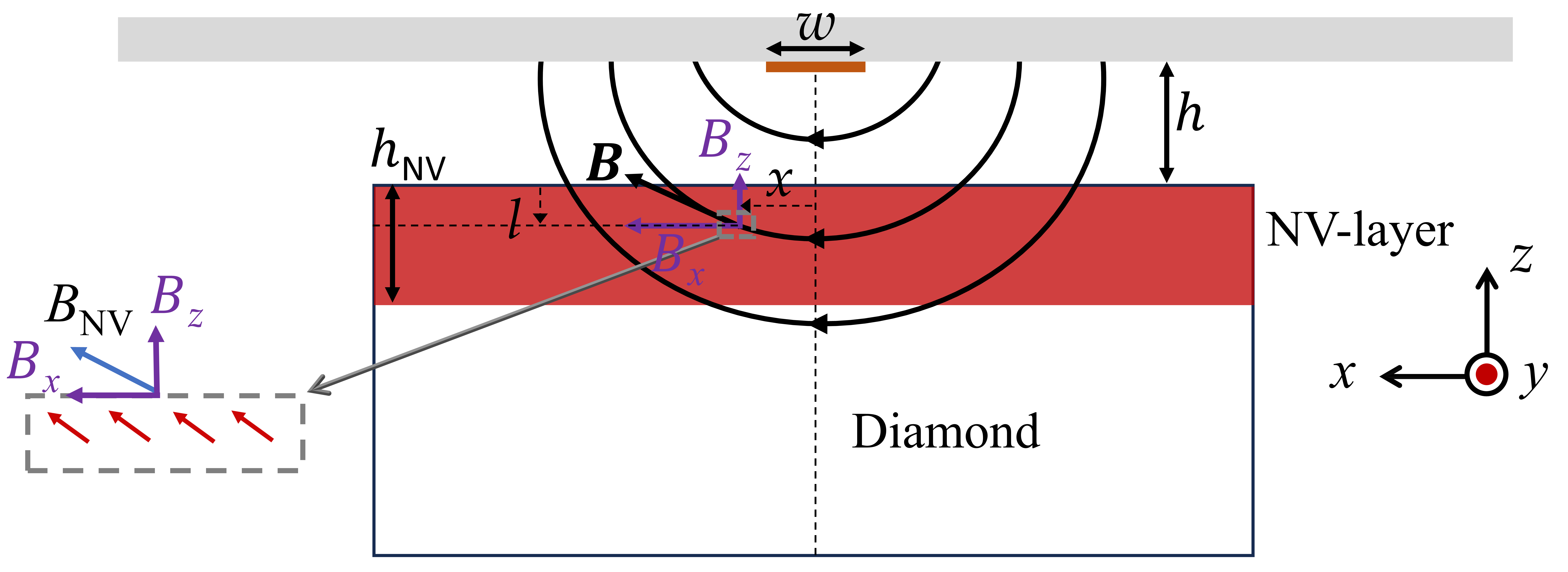}
    \caption{Schematic view of the magnetic field $\vb*{B}$ due to a current carrying 2D strip of width $w$, positioned at a stand-off distance $h$ above the diamond substrate containing an NV layer of thickness $h_{\text{NV}}$. The current is assumed to flow along $+y$-direction. $\vb*{B}$ is depicted at a lateral distance $x$ from the center of the strip and at a depth $l$ from the diamond surface. 
    The inset shows the projection of $\vb*{B}$ onto a single NV axis, corresponding to the measured magnetic field component $B_{\text{NV}}$.}
    \label{fig:stand_off}
\end{figure}

\hspace*{10pt} \textbf{Current Extraction in CM:  From Simulations and Conventional Method.}\label{Spice, conventional} The CM circuit is simulated using the Cadence Virtuoso tools with the a-IGZO TFT PDK. From the simulation, the circuit shows a current of  750 $\mu$A ($\mathcal{A}$) and 300 $\mu$A ($\mathcal{B}$) when the input current is 50 $\mu$A. For the same input, $\mathcal{A}$ shows 746.1 $\mu$A from conventional measurement using an edge card connector. Input current is supplied with  Keithley Model 6221 current source, and the output current is measured using  Keithley DMM6500 digital multimeter at a $V_{\text{DD}}$ of 2 V.
Further,  conventional approach allows access to only the pad area, and other conductive paths in the circuit cannot be accessed, thereby imposing a limitation on current extraction at different paths in the circuit.
\\
\hspace*{10pt} \textbf{Experimental Setup.} A home-built QDM setup is used for sensing and imaging the current flowing through the DUT. The 532 nm excitation laser beam reflects off a 550 nm cut-on long-pass dichroic mirror (Thorlabs DMLP550 Ø$2''$), passes through a plano-convex lens ($f$ = 400 mm), and is then incident on the back aperture of the air objective (Nikon LWD 40$\cross$ 0.55 NA). The diamond substrate is placed directly on the DUT, ensuring that the surface containing the NV-doped layer makes direct contact with the conducting circuit. The sensor assembly is positioned using a piezo-based nano-positioning stage integrated with a manually controlled micro-positioning system. This enables us to achieve control over the illumination of the circuit’s region of interest. The red fluorescence emitted from the NV layer is collected by the same objective, transmitted through the longpass dichroic mirror, filtered using a long-pass filter (Chroma ET655LP), and then focused onto the sCMOS camera (Andor Zyla 5.5) via a tube lens ($f$ = 250 mm). The effective magnification  of the objective and tube lens yields an effective pixel size of 130$\cross$130 nm$^2$. However, all measurements are performed using pixel binning, resulting in an effective pixel size of 260$\cross$260 nm$^2$. 
\\
\hspace*{10pt} Microwaves are generated using a signal generator (Keysight N5171B), amplified by a MW amplifier (Mini-Circuits ZHL-16W-43-S+), and delivered through a microwave circulator (DiTom D3C2040) to a custom-made MW loop antenna. This loop consists of a copper wire (1 mm diameter) soldered to an SMA connector and is positioned immediately above and close to the diamond-DUT interface. A bias magnetic field of 2-3 mT is applied using an NdFeB magnet. The magnetic field is aligned such that it orients along the $[111]$ crystallographic direction. Two separate sets of ODMR measurement data are acquired to determine the magnetic field. One set corresponds to the NV spin transition frequency $f_-$, while the other corresponds to the NV spin transition frequency $f_+$. To further enhance ODMR contrast, a half-wave plate (Thorlabs AHWP10M-600) is placed in the excitation path upstream of the focusing lens ($f$ = 400 mm). 
\\ 
\hspace*{10pt} \textbf{ODMR Data Acquisition.} CW-ODMR measurements involve application of laser excitation while sweeping the frequency of the microwaves to probe the spin state population of the diamond NV centers. Each MW frequency sweep, centered around the resonance frequency $f_-$ ($f_+$) consists of 71 discrete frequency points with a dwell time of 10 ms. At each frequency point, PL images are captured by the camera with an exposure time of 10 ms. To improve the SNR of the ODMR spectrum, MW frequency sweeps are repeated $\sim$110 times, and the resulting fluorescence images are averaged together. The entire measurement sequence takes roughly 10 minutes, yielding two distinct ODMR datasets corresponding to the resonances $f_-$ and $f_+$. Each ODMR dataset also features a two-fold splitting of the ODMR transitions due to the hyperfine coupling between the NV spin and the $^{15}$N nuclear spin. Following the methodology outlined in Reference~\cite{glenn}, individual ODMR spectra are extracted at a pixel level and subsequently analyzed using a two-dip Lorentzian fitting function ($\mathcal{L}(f)$) to account for the hyperfine splitting. The fitting function is explicitly defined as: 
\\
\begin{align}
\mathcal{L}(f) = \sum_{j=-,+} \Bigg[ 1 
    & - \frac{c^j_1}{1 + \left( \frac{f - f^j_1 }{\Delta^j_1} \right)^2} \nonumber \\
    & - \frac{c^j_2}{1 + \left( \frac{f - f^j_2 }{\Delta^j_2} \right)^2} 
\Bigg] + C,
\label{eq:BNV}
\end{align}
where the parameters $c^j_{1,2}$, $\Delta^j_{1,2}$, $f^j_{1,2}$, and $C$ represent the amplitude, linewidth, resonance frequencies, and background offset of the Lorentzian lineshapes, respectively. 
For each pixel, the resonance frequencies $f_-$ and $f_+$ are calculated by averaging the fitted hyperfine peak frequencies ($f^j_{1,2}$).
Finally, the magnetic field at each pixel position is determined from the frequency difference using the relation $B_{\text{NV}} = \frac{f_{+}-f_{-}}{2\gamma_{\text{NV}}}$, resulting in a two-dimensional magnetic field map.\\\\
\hspace*{10pt} \textbf{Determination of Stand-off Distance.} Using Biot-Savart law for a 2D conducting strip carrying a current $\mathcal{I}$, the magnetic field generated can be described by the following equations\cite{Tetienne:2019} (\figref{fig:stand_off}): 
\begin{subequations}\label{eq:B_fields}
\begin{align}
B_z &= -\frac{1}{2} \left(\frac{\mu_0 \mathcal{I}}{2 \pi w}\right)
\ln\left(\frac{\left(l+h\right)^2 + \left(x + \frac{w}{2}\right)^2}
{\left(l+h\right)^2 + \left(x - \frac{w}{2}\right)^2}\right) \label{eq:Bz}, \\
B_x &= \frac{\mu_0 \mathcal{I}}{2 \pi w} 
\left[\tan^{-1}\left(\frac{x - \frac{w}{2}}{l+h}\right) - 
\tan^{-1}\left(\frac{x + \frac{w}{2}}{l+h}\right)\right], \label{eq:By}
\end{align}
\end{subequations}
where \(w\) is the width of the wire and $\mu_0 = 4\pi \times 10^{-7} \, \text{m} \cdot \text{T} \cdot \text{A}^{-1}$.  
The above equations are valid under the assumption thickness $(t) \ll w$. Since $t = 0.3$\space$\mu$m, $w_{\mathcal{A}} = 9.5$\space$\mu$m and $w_{\mathcal{B}} = 15.5$\space$\mu$m, the equations can be applied to our case.
To account for the finite NV layer thickness, the measured $B_{\text{NV}}$ is corrected by averaging it over $h_\text{NV}$ (= 2.5 $\mu$m),\cite{Abrahams_2021}
\begin{equation}
B^\prime_{\text{NV}} = \frac{1}{h_{\text{NV}}} \int_{0}^{h_{\text{NV}}} B_{\text{NV}}(x, l) \, dl.
\label{eq:BNV_prime}
\end{equation}
The stand-off distances  $h_{\mathcal{A}} = 2.3$\space$\mu$m and $h_{\mathcal{B}} = 2.8$\space$\mu$m are thus determined by fitting the experimental data in \figref{fig:mf_image} to Eq. \ref{eq:BNV_prime}.
\section*{ACKNOWLEDGMENTS}
This work was conducted at the Indian Institute of Science Education and Research, Bhopal. M.Y.A.K. gratefully acknowledges financial support from the Chanakya PhD fellowship, provided by the National Mission on Interdisciplinary Cyber-Physical Systems, Department of Science and Technology, Government of India, through the I-HUB Quantum Technology Foundation Grant Number I-HUB/DF/2021-22/004. P.G.B. gratefully acknowledges financial support from the STARS MoE Grant Number MoE-STARS/STARS-2/2023-0327 and the India-Portugal Bilateral Cooperation Grant Number DST/INT/PORTUGAL/P11/2023(c). P.P. gratefully acknowledges financial support from the I-HUB Quantum Technology Foundation Grant Number I-HUB/SPOKE/2023-24/003 and STARS MoE Grant Number MoE-STARS/STARS1/662. Circuit fabrication was performed at the National Centre of Excellence for Large Area Flexible Electronics (NCFlexE), funded by the Ministry of Electronics and Information Technology, Government of India, at IIT Kanpur. The authors also acknowledge the Semiconductor Laboratory (SCL), Department of Space, Government of India, Sector 72, Mohali, for providing dicing services.
\section*{AUTHOR INFORMATION}
\subsection*{\textbf{Corresponding Authors}}
\noindent \textbf{Phani Kumar Peddibhotla} - Department of Physics, Indian Institute of Science Education and Research, Bhopal, India; https://orcid.org/0000-0002-0340-0769; Email: phani@iiserb.ac.in \\
\textbf{Pydi Ganga Bahubalindruni} - Department of Electrical Engineering and Computer Science, Indian Institute of Science Education and Research, Bhopal, India;  https://orcid.org/0000-0002-6241-5055; Email: ganga@iiserb.ac.in
\subsection*{\textbf{Author Contributions}}
\noindent M.Y.A.K. and P.D. contributed equally to this work. The manuscript was written through contributions of all authors. All authors have given approval to the final version of the manuscript.
\subsection*{\textbf{Conflict of Interest}}
\noindent The authors have no conflicts to disclose.
\bibliography{cite}

\providecommand{\latin}[1]{#1}
\makeatletter
\providecommand{\doi}
  {\begingroup\let\do\@makeother\dospecials
  \catcode`\{=1 \catcode`\}=2 \doi@aux}
\providecommand{\doi@aux}[1]{\endgroup\texttt{#1}}
\makeatother
\providecommand*\mcitethebibliography{\thebibliography}
\csname @ifundefined\endcsname{endmcitethebibliography}  {\let\endmcitethebibliography\endthebibliography}{}
\begin{mcitethebibliography}{51}
\providecommand*\natexlab[1]{#1}
\providecommand*\mciteSetBstSublistMode[1]{}
\providecommand*\mciteSetBstMaxWidthForm[2]{}
\providecommand*\mciteBstWouldAddEndPuncttrue
  {\def\EndOfBibitem{\unskip.}}
\providecommand*\mciteBstWouldAddEndPunctfalse
  {\let\EndOfBibitem\relax}
\providecommand*\mciteSetBstMidEndSepPunct[3]{}
\providecommand*\mciteSetBstSublistLabelBeginEnd[3]{}
\providecommand*\EndOfBibitem{}
\mciteSetBstSublistMode{f}
\mciteSetBstMaxWidthForm{subitem}{(\alph{mcitesubitemcount})}
\mciteSetBstSublistLabelBeginEnd
  {\mcitemaxwidthsubitemform\space}
  {\relax}
  {\relax}

\bibitem[Knauss \latin{et~al.}(2001)Knauss, Cawthorne, Lettsome, Kelly, Chatraphorn, Fleet, Wellstood, and Vanderlinde]{knauss2001scanning}
Knauss,~L.~A.; Cawthorne,~A.; Lettsome,~N.; Kelly,~S.; Chatraphorn,~S.; Fleet,~E.; Wellstood,~F.; Vanderlinde,~W. \href{https://doi.org/10.1016/S0026-2714(01)00108-1}{Scanning SQUID microscopy for current imaging}. \emph{Microelectronics Reliability} \textbf{2001}, \emph{41}, 1211--1229\relax
\mciteBstWouldAddEndPuncttrue
\mciteSetBstMidEndSepPunct{\mcitedefaultmidpunct}
{\mcitedefaultendpunct}{\mcitedefaultseppunct}\relax
\EndOfBibitem
\bibitem[Fong \latin{et~al.}(2005)Fong, Holzer, McBride, Lima, Baudenbacher, and Radparvar]{fong}
Fong,~L.~E.; Holzer,~J.~R.; McBride,~K.~K.; Lima,~E.~A.; Baudenbacher,~F.; Radparvar,~M. \href{https://doi.org/10.1063/1.1884025}{High-resolution room-temperature sample scanning superconducting quantum interference device microscope configurable for geological and biomagnetic applications}. \emph{Rev. Sci. Instrum.} \textbf{2005}, \emph{76}\relax
\mciteBstWouldAddEndPuncttrue
\mciteSetBstMidEndSepPunct{\mcitedefaultmidpunct}
{\mcitedefaultendpunct}{\mcitedefaultseppunct}\relax
\EndOfBibitem
\bibitem[Dilorio \latin{et~al.}(1991)Dilorio, Yoshizumi, Maung, Yang, Zhang, and Fan]{dilorio1991manufacturable}
Dilorio,~M.; Yoshizumi,~S.; Maung,~M.; Yang,~K.-Y.; Zhang,~J.; Fan,~N. \href{https://doi.org/10.1038/354513a0}{Manufacturable low-noise SQUIDs operating in liquid nitrogen}. \emph{Nature} \textbf{1991}, \emph{354}, 513--515\relax
\mciteBstWouldAddEndPuncttrue
\mciteSetBstMidEndSepPunct{\mcitedefaultmidpunct}
{\mcitedefaultendpunct}{\mcitedefaultseppunct}\relax
\EndOfBibitem
\bibitem[Horn \latin{et~al.}(2019)Horn, Cui, Kirtley, and Moler]{horn2019cryogen}
Horn,~B.-V.; Cui,~Z.; Kirtley,~J.~R.; Moler,~K.~A. \href{https://doi.org/10.1063/1.5085008}{Cryogen-free variable temperature scanning SQUID microscope}. \emph{Rev. Sci. Instrum.} \textbf{2019}, \emph{90}\relax
\mciteBstWouldAddEndPuncttrue
\mciteSetBstMidEndSepPunct{\mcitedefaultmidpunct}
{\mcitedefaultendpunct}{\mcitedefaultseppunct}\relax
\EndOfBibitem
\bibitem[Herrera-May \latin{et~al.}(2009)Herrera-May, Aguilera-Cortés, García-Ramírez, and Manjarrez]{Herrera_2009}
Herrera-May,~A.~L.; Aguilera-Cortés,~L.~A.; García-Ramírez,~P.~J.; Manjarrez,~E. \href{https://www.mdpi.com/1424-8220/9/10/7785}{Resonant Magnetic Field Sensors Based On MEMS Technology}. \emph{Sensors} \textbf{2009}, \emph{9}, 7785--7813\relax
\mciteBstWouldAddEndPuncttrue
\mciteSetBstMidEndSepPunct{\mcitedefaultmidpunct}
{\mcitedefaultendpunct}{\mcitedefaultseppunct}\relax
\EndOfBibitem
\bibitem[Levine \latin{et~al.}(2019)Levine, Turner, Kehayias, Hart, Langellier, Trubko, Glenn, and Walsworth]{levine}
Levine,~E.~V.; Turner,~M.~J.; Kehayias,~P.; Hart,~C.~A.; Langellier,~N.; Trubko,~R.; Glenn,~D.~R.; Walsworth,~R.~L. \href{https://doi.org/10.1515/nanoph-2019-0209}{Principles and techniques of the quantum diamond microscope}. \emph{Nanophotonics} \textbf{2019}, \emph{8}, 1945\relax
\mciteBstWouldAddEndPuncttrue
\mciteSetBstMidEndSepPunct{\mcitedefaultmidpunct}
{\mcitedefaultendpunct}{\mcitedefaultseppunct}\relax
\EndOfBibitem
\bibitem[Glenn \latin{et~al.}(2017)Glenn, Fu, Kehayias, Le~Sage, Lima, Weiss, and Walsworth]{glenn}
Glenn,~D.~R.; Fu,~R.~R.; Kehayias,~P.; Le~Sage,~D.; Lima,~E.~A.; Weiss,~B.~P.; Walsworth,~R.~L. \href{https://doi.org/10.1002/2017GC006946}{Micrometer-scale magnetic imaging of geological samples using a quantum diamond microscope}. \emph{Geochem. Geophys. Geosyst.} \textbf{2017}, \emph{18}, 3254\relax
\mciteBstWouldAddEndPuncttrue
\mciteSetBstMidEndSepPunct{\mcitedefaultmidpunct}
{\mcitedefaultendpunct}{\mcitedefaultseppunct}\relax
\EndOfBibitem
\bibitem[Scholten \latin{et~al.}(2021)Scholten, Healey, Robertson, Abrahams, Broadway, and Tetienne]{Scholten_2021}
Scholten,~S.~C.; Healey,~A.~J.; Robertson,~I.~O.; Abrahams,~G.~J.; Broadway,~D.~A.; Tetienne,~J.-P. \href{https://doi.org/10.1063/5.0066733}{Widefield quantum microscopy with nitrogen-vacancy centers in diamond: strengths, limitations, and prospects}. \emph{J. Appl. Phys.} \textbf{2021}, \emph{130}, 150902\relax
\mciteBstWouldAddEndPuncttrue
\mciteSetBstMidEndSepPunct{\mcitedefaultmidpunct}
{\mcitedefaultendpunct}{\mcitedefaultseppunct}\relax
\EndOfBibitem
\bibitem[Hudak and Stroud(2023)Hudak, and Stroud]{hudak2023atomically}
Hudak,~B.~M.; Stroud,~R.~M. \href{https://pubs.acs.org/doi/10.1021/acsnano.2c10122}{Atomically Precise Detection and Manipulation of Nitrogen-Vacancy Centers in Nanodiamonds}. \emph{ACS Nano} \textbf{2023}, \emph{17}, 7241--7249\relax
\mciteBstWouldAddEndPuncttrue
\mciteSetBstMidEndSepPunct{\mcitedefaultmidpunct}
{\mcitedefaultendpunct}{\mcitedefaultseppunct}\relax
\EndOfBibitem
\bibitem[Xu \latin{et~al.}(2025)Xu, Palm, Huxter, Herb, Abendroth, Bouzehouane, Boulle, Gabor, Urrestarazu~Larranaga, Morales, \latin{et~al.} others]{xu2025minimizing}
Xu,~Z.; Palm,~M.~L.; Huxter,~W.; Herb,~K.; Abendroth,~J.~M.; Bouzehouane,~K.; Boulle,~O.; Gabor,~M.~S.; Urrestarazu~Larranaga,~J.; Morales,~A.; others \href{https://pubs.acs.org/doi/10.1021/acsnano.4c18460}{Minimizing sensor-sample distances in scanning nitrogen-vacancy magnetometry}. \emph{ACS Nano} \textbf{2025}, \emph{19}, 8255--8265\relax
\mciteBstWouldAddEndPuncttrue
\mciteSetBstMidEndSepPunct{\mcitedefaultmidpunct}
{\mcitedefaultendpunct}{\mcitedefaultseppunct}\relax
\EndOfBibitem
\bibitem[Kehayias \latin{et~al.}(2022)Kehayias, Levine, Basso, Henshaw, Saleh~Ziabari, Titze, Haltli, Okoro, Tibbetts, Udoni, Bielejec, Lilly, Lu, Schwindt, and Mounce]{Kehayias:2022}
Kehayias,~P.; Levine,~E.~V.; Basso,~L.; Henshaw,~J.; Saleh~Ziabari,~M.; Titze,~M.; Haltli,~R.; Okoro,~J.; Tibbetts,~D.~R.; Udoni,~D.~M.; Bielejec,~E.; Lilly,~M.~P.; Lu,~T.-M.; Schwindt,~P. D.~D.; Mounce,~A.~M. \href{https://link.aps.org/doi/10.1103/PhysRevApplied.17.014021}{Measurement and Simulation of the Magnetic Fields from a 555 Timer Integrated Circuit Using a Quantum Diamond Microscope and Finite-Element Analysis}. \emph{Phys. Rev. Appl.} \textbf{2022}, \emph{17}, 014021\relax
\mciteBstWouldAddEndPuncttrue
\mciteSetBstMidEndSepPunct{\mcitedefaultmidpunct}
{\mcitedefaultendpunct}{\mcitedefaultseppunct}\relax
\EndOfBibitem
\bibitem[Basso \latin{et~al.}(2022)Basso, Kehayias, Henshaw, Saleh~Ziabari, Byeon, Lilly, Bussmann, Campbell, Misra, and Mounce]{Basso:2023}
Basso,~L.; Kehayias,~P.; Henshaw,~J.; Saleh~Ziabari,~M.; Byeon,~H.; Lilly,~M.~P.; Bussmann,~E.; Campbell,~D.~M.; Misra,~S.; Mounce,~A.~M. \href{https://dx.doi.org/10.1088/1361-6528/ac95a0}{Electric current paths in a Si:P delta-doped device imaged by nitrogen-vacancy diamond magnetic microscopy}. \emph{Nanotechnology} \textbf{2022}, \emph{34}, 015001\relax
\mciteBstWouldAddEndPuncttrue
\mciteSetBstMidEndSepPunct{\mcitedefaultmidpunct}
{\mcitedefaultendpunct}{\mcitedefaultseppunct}\relax
\EndOfBibitem
\bibitem[Scholten \latin{et~al.}(2022)Scholten, Robertson, Abrahams, Singh, Healey, and Tetienne]{scholten2022imaging}
Scholten,~S.~C.; Robertson,~I.~O.; Abrahams,~G.~J.; Singh,~P.; Healey,~A.~J.; Tetienne,~J.-P. \href{https://doi.org/10.1116/5.0114436}{Aberration control in quantitative widefield quantum microscopy}. \emph{AVS Quantum Sci.} \textbf{2022}, \emph{4}, 034404\relax
\mciteBstWouldAddEndPuncttrue
\mciteSetBstMidEndSepPunct{\mcitedefaultmidpunct}
{\mcitedefaultendpunct}{\mcitedefaultseppunct}\relax
\EndOfBibitem
\bibitem[Tetienne \latin{et~al.}(2017)Tetienne, Dontschuk, Broadway, Stacey, Simpson, and Hollenberg]{Tetienne:2017}
Tetienne,~J.-P.; Dontschuk,~N.; Broadway,~D.~A.; Stacey,~A.; Simpson,~D.~A.; Hollenberg,~L. C.~L. \href{https://doi.org/10.1126/sciadv.1602429}{Quantum imaging of current flow in graphene}. \emph{Sci. Adv.} \textbf{2017}, \emph{3}, 1602429\relax
\mciteBstWouldAddEndPuncttrue
\mciteSetBstMidEndSepPunct{\mcitedefaultmidpunct}
{\mcitedefaultendpunct}{\mcitedefaultseppunct}\relax
\EndOfBibitem
\bibitem[Chang \latin{et~al.}(2017)Chang, Eichler, Rhensius, Lorenzelli, Maletinsky, and Degen]{chang}
Chang,~K.; Eichler,~A.; Rhensius,~J.; Lorenzelli,~L.; Maletinsky,~P.; Degen,~C.~L. \href{https://doi.org/10.1021/acs.nanolett.6b05304}{Nanoscale Imaging of Current Density with a Single-Spin Magnetometer}. \emph{Nano Lett.} \textbf{2017}, \emph{17}, 2367\relax
\mciteBstWouldAddEndPuncttrue
\mciteSetBstMidEndSepPunct{\mcitedefaultmidpunct}
{\mcitedefaultendpunct}{\mcitedefaultseppunct}\relax
\EndOfBibitem
\bibitem[Zhong \latin{et~al.}(2024)Zhong, Wang, Mai, Ye, Li, Wang, Dai, Wang, Sun, and Zhang]{Zhong_2024}
Zhong,~C.; Wang,~Y.; Mai,~D.; Ye,~C.; Li,~X.; Wang,~H.; Dai,~R.; Wang,~Z.; Sun,~X.; Zhang,~Z. \href{https://doi.org/10.1021/acs.nanolett.4c00780}{High Spatial Resolution 2D Imaging of Current Density and Pressure for Graphene Devices under High Pressure Using Nitrogen-Vacancy Centers in Diamond}. \emph{Nano Lett.} \textbf{2024}, \emph{24}, 4993\relax
\mciteBstWouldAddEndPuncttrue
\mciteSetBstMidEndSepPunct{\mcitedefaultmidpunct}
{\mcitedefaultendpunct}{\mcitedefaultseppunct}\relax
\EndOfBibitem
\bibitem[Garsi \latin{et~al.}(2024)Garsi, St{\"o}hr, Denisenko, Shagieva, Trautmann, Vogl, Sene, Kaiser, Zappe, Reuter, and Wrachtrup]{Garsi:2024}
Garsi,~M.; St{\"o}hr,~R.; Denisenko,~A.; Shagieva,~F.; Trautmann,~N.; Vogl,~U.; Sene,~B.; Kaiser,~F.; Zappe,~A.; Reuter,~R.; Wrachtrup,~J. \href{https://link.aps.org/doi/10.1103/PhysRevApplied.21.014055}{Three-dimensional imaging of integrated-circuit activity using quantum defects in diamond}. \emph{Phys. Rev. Appl.} \textbf{2024}, \emph{21}, 014055\relax
\mciteBstWouldAddEndPuncttrue
\mciteSetBstMidEndSepPunct{\mcitedefaultmidpunct}
{\mcitedefaultendpunct}{\mcitedefaultseppunct}\relax
\EndOfBibitem
\bibitem[Turner \latin{et~al.}(2020)Turner, Langellier, Bainbridge, Walters, Meesala, Babinec, Kehayias, Yacoby, Hu, Lon{\v{c}}ar, Walsworth, and Levine]{Turner:2020}
Turner,~M.~J.; Langellier,~N.; Bainbridge,~R.; Walters,~D.; Meesala,~S.; Babinec,~T.~M.; Kehayias,~P.; Yacoby,~A.; Hu,~E.; Lon{\v{c}}ar,~M.; Walsworth,~R.~L.; Levine,~E.~V. \href{https://doi.org/10.1103/PhysRevApplied.14.014097}{Magnetic Field Fingerprinting of Integrated-Circuit Activity with a Quantum Diamond Microscope}. \emph{Phys. Rev. Appl.} \textbf{2020}, \emph{14}, 014097\relax
\mciteBstWouldAddEndPuncttrue
\mciteSetBstMidEndSepPunct{\mcitedefaultmidpunct}
{\mcitedefaultendpunct}{\mcitedefaultseppunct}\relax
\EndOfBibitem
\bibitem[Oliver \latin{et~al.}(2022)Oliver, Martynowych, Turner, Hopper, Walsworth, and Levine]{Oliver}
Oliver,~S.~M.; Martynowych,~D.~J.; Turner,~M.~J.; Hopper,~D.~A.; Walsworth,~R.~L.; Levine,~E.~V. \href{https://doi.org/10.31399/asm.cp.istfa2021p0096}{Vector magnetic current imaging of an 8 nm process node chip and 3D current distributions using the quantum diamond microscope}. \emph{ISTFA} \textbf{2022}, 96--107\relax
\mciteBstWouldAddEndPuncttrue
\mciteSetBstMidEndSepPunct{\mcitedefaultmidpunct}
{\mcitedefaultendpunct}{\mcitedefaultseppunct}\relax
\EndOfBibitem
\bibitem[Nakano \latin{et~al.}(2012)Nakano, Saito, Miura, Sakano, Ueda, Sugi, Yamaguchi, Amemiya, Hiramatsu, and Ishida]{nakano}
Nakano,~S.; Saito,~N.; Miura,~K.; Sakano,~T.; Ueda,~T.; Sugi,~K.; Yamaguchi,~H.; Amemiya,~I.; Hiramatsu,~M.; Ishida,~A. \href{https://doi.org/10.1002/jsid.111}{Highly reliable a-IGZO TFTs on a plastic substrate for flexible AMOLED displays}. \emph{J. Soc. Inf. Disp.} \textbf{2012}, \emph{20}, 493--498\relax
\mciteBstWouldAddEndPuncttrue
\mciteSetBstMidEndSepPunct{\mcitedefaultmidpunct}
{\mcitedefaultendpunct}{\mcitedefaultseppunct}\relax
\EndOfBibitem
\bibitem[Zhang \latin{et~al.}(2022)Zhang, He, Wang, Wang, Xu, and Liu]{zhang2022ultraviolet}
Zhang,~Y.; He,~G.; Wang,~L.; Wang,~W.; Xu,~X.; Liu,~W. \href{https://pubs.acs.org/doi/10.1021/acsnano.2c01286}{Ultraviolet-Assisted Low-Thermal-Budget-Driven $\alpha$-InGaZnO Thin Films for High-Performance Transistors and Logic Circuits}. \emph{ACS Nano} \textbf{2022}, \emph{16}, 4961--4971\relax
\mciteBstWouldAddEndPuncttrue
\mciteSetBstMidEndSepPunct{\mcitedefaultmidpunct}
{\mcitedefaultendpunct}{\mcitedefaultseppunct}\relax
\EndOfBibitem
\bibitem[Kim \latin{et~al.}(2018)Kim, Byun, Jang, Kim, Han, Park, and Choi]{kim}
Kim,~J.-S.; Byun,~J.-W.; Jang,~J.-H.; Kim,~Y.-D.; Han,~K.-L.; Park,~J.-S.; Choi,~B.-D. \href{https://doi.org/10.1109/TED.2018.2843180}{A High-Reliability Carry-Free Gate Driver for Flexible Displays Using a-IGZO TFTs}. \emph{IEEE Trans. Electron Devices} \textbf{2018}, \emph{65}, 3269\relax
\mciteBstWouldAddEndPuncttrue
\mciteSetBstMidEndSepPunct{\mcitedefaultmidpunct}
{\mcitedefaultendpunct}{\mcitedefaultseppunct}\relax
\EndOfBibitem
\bibitem[Wieczorek \latin{et~al.}(2023)Wieczorek, Starecki, Go{\l}ofit, Radtke, and Pilarz]{wieczorek2023thin}
Wieczorek,~P.~Z.; Starecki,~K.; Go{\l}ofit,~K.; Radtke,~M.; Pilarz,~M. \href{https://ieeexplore.ieee.org/document/10214492}{A thin elastic NFC Forum type 1 compatible RFID tag}. \emph{Solid-State Circuits} \textbf{2023}, \emph{59}, 935--946\relax
\mciteBstWouldAddEndPuncttrue
\mciteSetBstMidEndSepPunct{\mcitedefaultmidpunct}
{\mcitedefaultendpunct}{\mcitedefaultseppunct}\relax
\EndOfBibitem
\bibitem[Meister \latin{et~al.}(2020)Meister, Ishida, Sou, Carta, and Ellinger]{meister202049}
Meister,~T.; Ishida,~K.; Sou,~A.; Carta,~C.; Ellinger,~F. \href{https://doi.org/10.1049/el.2020.0813}{49.35 MHz GBW and 33.43 MHz GBW amplifiers in flexible a-IGZO TFT technology}. \emph{Electronics Letters} \textbf{2020}, \emph{56}, 782--785\relax
\mciteBstWouldAddEndPuncttrue
\mciteSetBstMidEndSepPunct{\mcitedefaultmidpunct}
{\mcitedefaultendpunct}{\mcitedefaultseppunct}\relax
\EndOfBibitem
\bibitem[Monga and Halonen(2024)Monga, and Halonen]{monga2024flexible}
Monga,~D.~C.; Halonen,~K. \href{https://ieeexplore.ieee.org/document/10603873}{Flexible RF to DC Converter for Wireless Power Transfer in NFC and Biomedical Systems}. \emph{FLEPS} \textbf{2024}, 1--4\relax
\mciteBstWouldAddEndPuncttrue
\mciteSetBstMidEndSepPunct{\mcitedefaultmidpunct}
{\mcitedefaultendpunct}{\mcitedefaultseppunct}\relax
\EndOfBibitem
\bibitem[Geng \latin{et~al.}(2023)Geng, Wang, Li, Myny, Nathan, Jang, Kuo, and Liu]{geng2023thin}
Geng,~D.; Wang,~K.; Li,~L.; Myny,~K.; Nathan,~A.; Jang,~J.; Kuo,~Y.; Liu,~M. \href{https://doi.org/10.1038/s41928-023-01095-8}{Thin-film transistors for large-area electronics}. \emph{Nature Electronics} \textbf{2023}, \emph{6}, 963--972\relax
\mciteBstWouldAddEndPuncttrue
\mciteSetBstMidEndSepPunct{\mcitedefaultmidpunct}
{\mcitedefaultendpunct}{\mcitedefaultseppunct}\relax
\EndOfBibitem
\bibitem[Zhao \latin{et~al.}(2024)Zhao, Li, Liu, Li, Chen, and Wu]{zhao}
Zhao,~M.; Li,~L.; Liu,~R.; Li,~B.; Chen,~R.; Wu,~Z. \href{https://ieeexplore.ieee.org/document/10518128}{A 43.5dB Gain Unipolar a-IGZO TFT Amplifier with Parallel Bootstrap Capacitor for Bio-signals Sensing Applications}. \emph{IEEE Trans. Biomed. Circuits Syst.} \textbf{2024}, \emph{18}, 1371\relax
\mciteBstWouldAddEndPuncttrue
\mciteSetBstMidEndSepPunct{\mcitedefaultmidpunct}
{\mcitedefaultendpunct}{\mcitedefaultseppunct}\relax
\EndOfBibitem
\bibitem[Shrivastava \latin{et~al.}(2023)Shrivastava, Bahubalindruni, and Choudhary]{shri}
Shrivastava,~S.; Bahubalindruni,~P.~G.; Choudhary,~V. \href{https://doi.org/10.1109/JFLEX.2023.3304598}{Temperature Detection System Using Oxide TFTs on a Flexible Substrate}. \emph{IEEE J. Flex. Electron.} \textbf{2023}, \emph{2}, 464\relax
\mciteBstWouldAddEndPuncttrue
\mciteSetBstMidEndSepPunct{\mcitedefaultmidpunct}
{\mcitedefaultendpunct}{\mcitedefaultseppunct}\relax
\EndOfBibitem
\bibitem[Velazquez~Lopez \latin{et~al.}(2024)Velazquez~Lopez, Linares-Barranco, Lee, Erfanijazi, Patino-Saucedo, Sifalakis, Catthoor, and Myny]{velazquez2024tunable}
Velazquez~Lopez,~M.; Linares-Barranco,~B.; Lee,~J.; Erfanijazi,~H.; Patino-Saucedo,~A.; Sifalakis,~M.; Catthoor,~F.; Myny,~K. \href{https://doi.org/10.1038/s44172-024-00248-7}{A tunable multi-timescale Indium-Gallium-Zinc-Oxide thin-film transistor neuron towards hybrid solutions for spiking neuromorphic applications}. \emph{Nature Commun. Eng.} \textbf{2024}, \emph{3}, 102\relax
\mciteBstWouldAddEndPuncttrue
\mciteSetBstMidEndSepPunct{\mcitedefaultmidpunct}
{\mcitedefaultendpunct}{\mcitedefaultseppunct}\relax
\EndOfBibitem
\bibitem[Jin \latin{et~al.}(2015)Jin, Kang, Cho, Han, Chung, Lee, Shin, Baek, Kim, Lee, \latin{et~al.} others]{jin2015water}
Jin,~S.~H.; Kang,~S.-K.; Cho,~I.-T.; Han,~S.~Y.; Chung,~H.~U.; Lee,~D.~J.; Shin,~J.; Baek,~G.~W.; Kim,~T.-i.; Lee,~J.-H.; others \href{https://pubs.acs.org/doi/10.1021/acsami.5b00086}{Water-soluble thin film transistors and circuits based on amorphous indium--gallium--zinc oxide}. \emph{ACS applied materials \& interfaces} \textbf{2015}, \emph{7}, 8268--8274\relax
\mciteBstWouldAddEndPuncttrue
\mciteSetBstMidEndSepPunct{\mcitedefaultmidpunct}
{\mcitedefaultendpunct}{\mcitedefaultseppunct}\relax
\EndOfBibitem
\bibitem[Jang \latin{et~al.}(2022)Jang, Park, Kang, and Lee]{jang2022amorphous}
Jang,~Y.; Park,~J.; Kang,~J.; Lee,~S.-Y. \href{https://doi.org/10.1021/acsaelm.1c01088}{Amorphous InGaZnO (a-IGZO) synaptic transistor for neuromorphic computing}. \emph{ACS Applied Electronic Materials} \textbf{2022}, \emph{4}, 1427--1448\relax
\mciteBstWouldAddEndPuncttrue
\mciteSetBstMidEndSepPunct{\mcitedefaultmidpunct}
{\mcitedefaultendpunct}{\mcitedefaultseppunct}\relax
\EndOfBibitem
\bibitem[Kim \latin{et~al.}(2021)Kim, Lee, Kang, Nam, Kim, Kim, and Park]{kim2021stress}
Kim,~K.-T.; Lee,~K.~W.; Kang,~S.-H.; Nam,~S.-J.; Kim,~J.; Kim,~Y.-H.; Park,~S.~K. \href{https://pubs.acs.org/doi/10.1021/acsaelm.1c00751}{Stress-released amorphous oxide/carbon nanotube CMOS amplifier circuits for skin-compatible electronics}. \emph{ACS Applied Electronic Materials} \textbf{2021}, \emph{3}, 4950--4958\relax
\mciteBstWouldAddEndPuncttrue
\mciteSetBstMidEndSepPunct{\mcitedefaultmidpunct}
{\mcitedefaultendpunct}{\mcitedefaultseppunct}\relax
\EndOfBibitem
\bibitem[Jia \latin{et~al.}(2023)Jia, Zhang, Liu, Li, Wang, Zhong, Han, Qin, and Huang]{jia}
Jia,~B.; Zhang,~C.; Liu,~M.; Li,~Z.; Wang,~J.; Zhong,~L.; Han,~C.; Qin,~M.; Huang,~X. \href{https://www.nature.com/articles/s41467-023-41181-1}{Integration of microbattery with thin-film electronics for constructing an integrated transparent microsystem based on InGaZnO}. \emph{Nat. Commun.} \textbf{2023}, \emph{14}, 5330\relax
\mciteBstWouldAddEndPuncttrue
\mciteSetBstMidEndSepPunct{\mcitedefaultmidpunct}
{\mcitedefaultendpunct}{\mcitedefaultseppunct}\relax
\EndOfBibitem
\bibitem[Qi \latin{et~al.}(2025)Qi, Li, Xia, Liu, Dai, Song, Yang, Ma, and Wang]{qi2025integrated}
Qi,~S.~M.; Li,~J.~C.; Xia,~Y.~H.; Liu,~Z.~C.; Dai,~D.; Song,~T.~L.; Yang,~H.~X.; Ma,~Y.~X.; Wang,~Y.~L. \href{https://pubs.acs.org/doi/10.1021/acsphotonics.4c01809}{Integrated Opto-Synaptic IGZO Transistors for Image Recognition Fabricated at Room Temperature}. \emph{ACS Photonics} \textbf{2025}, \relax
\mciteBstWouldAddEndPunctfalse
\mciteSetBstMidEndSepPunct{\mcitedefaultmidpunct}
{}{\mcitedefaultseppunct}\relax
\EndOfBibitem
\bibitem[Sen \latin{et~al.}(2022)Sen, Park, Pujar, Bala, Cho, Liu, Gandla, and Kim]{sen2022probing}
Sen,~A.; Park,~H.; Pujar,~P.; Bala,~A.; Cho,~H.; Liu,~N.; Gandla,~S.; Kim,~S. \href{https://pubs.acs.org/doi/10.1021/acsnano.2c08264}{Probing the Efficacy of Large-Scale Nonporous IGZO for Visible-to-NIR Detection Capability: An Approach toward High-Performance Image Sensor Circuitry}. \emph{ACS Nano} \textbf{2022}, \emph{16}, 9267--9277\relax
\mciteBstWouldAddEndPuncttrue
\mciteSetBstMidEndSepPunct{\mcitedefaultmidpunct}
{\mcitedefaultendpunct}{\mcitedefaultseppunct}\relax
\EndOfBibitem
\bibitem[Kim \latin{et~al.}(2018)Kim, Ahn, Jung, Cho, and Cho]{kim2018toward}
Kim,~K.~S.; Ahn,~C.~H.; Jung,~S.~H.; Cho,~S.~W.; Cho,~H.~K. \href{https://pubs.acs.org/doi/10.1021/acsami.7b18657}{Toward adequate operation of amorphous oxide thin-film transistors for low-concentration gas detection}. \emph{ACS applied materials \& interfaces} \textbf{2018}, \emph{10}, 10185--10193\relax
\mciteBstWouldAddEndPuncttrue
\mciteSetBstMidEndSepPunct{\mcitedefaultmidpunct}
{\mcitedefaultendpunct}{\mcitedefaultseppunct}\relax
\EndOfBibitem
\bibitem[Chu \latin{et~al.}(2022)Chu, Tan, Zhao, Wu, and Ding]{chu2022power}
Chu,~Y.; Tan,~H.; Zhao,~C.; Wu,~X.; Ding,~S.-J. \href{https://pubs.acs.org/doi/10.1021/acsami.1c19771}{Power-efficient gas-sensing and synaptic diodes based on lateral pentacene/a-IGZO pn junctions}. \emph{ACS Applied Materials \& Interfaces} \textbf{2022}, \emph{14}, 9368--9376\relax
\mciteBstWouldAddEndPuncttrue
\mciteSetBstMidEndSepPunct{\mcitedefaultmidpunct}
{\mcitedefaultendpunct}{\mcitedefaultseppunct}\relax
\EndOfBibitem
\bibitem[Nomura \latin{et~al.}(2004)Nomura, Ohta, Takagi, Kamiya, Hirano, and Hosono]{nomura2004room}
Nomura,~K.; Ohta,~H.; Takagi,~A.; Kamiya,~T.; Hirano,~M.; Hosono,~H. \href{https://www.nature.com/articles/nature03090}{Room-temperature fabrication of transparent flexible thin-film transistors using amorphous oxide semiconductors}. \emph{Nature} \textbf{2004}, \emph{432}, 488\relax
\mciteBstWouldAddEndPuncttrue
\mciteSetBstMidEndSepPunct{\mcitedefaultmidpunct}
{\mcitedefaultendpunct}{\mcitedefaultseppunct}\relax
\EndOfBibitem
\bibitem[Reuss \latin{et~al.}(2005)Reuss, Chalamala, Moussessian, Kane, Kumar, Zhang, Rogers, Hatalis, Temple, Moddel, Eliasson, Estes, Kunze, Handy, Harmon, Salzman, Woodall, Alam, Murthy, Jacobsen, Olivier, Markus, Campbell, and Snow]{reuss2005macroelectronics}
Reuss,~R.~H. \latin{et~al.}  \href{https://ieeexplore.ieee.org/document/1461580}{Macroelectronics: Perspectives on Technology and Applications}. \emph{Proc. IEEE} \textbf{2005}, \emph{93}, 1239\relax
\mciteBstWouldAddEndPuncttrue
\mciteSetBstMidEndSepPunct{\mcitedefaultmidpunct}
{\mcitedefaultendpunct}{\mcitedefaultseppunct}\relax
\EndOfBibitem
\bibitem[Han \latin{et~al.}(2021)Han, Lee, Kim, Kim, Choi, and Park]{han2021mechanical}
Han,~K.-L.; Lee,~W.-B.; Kim,~Y.-D.; Kim,~J.-H.; Choi,~B.-D.; Park,~J.-S. \href{https://doi.org/10.1021/acsaelm.1c00806}{Mechanical durability of flexible/stretchable a-IGZO TFTs on PI island for wearable electronic application}. \emph{ACS Applied Electronic Materials} \textbf{2021}, \emph{3}, 5037--5047\relax
\mciteBstWouldAddEndPuncttrue
\mciteSetBstMidEndSepPunct{\mcitedefaultmidpunct}
{\mcitedefaultendpunct}{\mcitedefaultseppunct}\relax
\EndOfBibitem
\bibitem[Shi \latin{et~al.}(2024)Shi, Tsuji, Cho, Ueda, Kim, and Hosono]{shi2024approach}
Shi,~Y.; Tsuji,~M.; Cho,~H.; Ueda,~S.; Kim,~J.; Hosono,~H. \href{https://pubs.acs.org/doi/10.1021/acsnano.4c02101}{Approach to Low Contact Resistance Formation on Buried Interface in Oxide Thin-Film Transistors: Utilization of Palladium-Mediated Hydrogen Pathway}. \emph{ACS Nano} \textbf{2024}, \emph{18}, 9736--9745\relax
\mciteBstWouldAddEndPuncttrue
\mciteSetBstMidEndSepPunct{\mcitedefaultmidpunct}
{\mcitedefaultendpunct}{\mcitedefaultseppunct}\relax
\EndOfBibitem
\bibitem[Troughton and Atkinson(2019)Troughton, and Atkinson]{troughton2019amorphous}
Troughton,~J.; Atkinson,~D. \href{http://dx.doi.org/10.1039/C9TC03933C}{Amorphous InGaZnO and metal oxide semiconductor devices: An overview and current status}. \emph{Journal of Materials Chemistry} \textbf{2019}, \emph{7}, 12388--12414\relax
\mciteBstWouldAddEndPuncttrue
\mciteSetBstMidEndSepPunct{\mcitedefaultmidpunct}
{\mcitedefaultendpunct}{\mcitedefaultseppunct}\relax
\EndOfBibitem
\bibitem[Bahubalindruni \latin{et~al.}(2013)Bahubalindruni, Barquinha, Duarte, de~Oliveira, Martins, Fortunato, \latin{et~al.} others]{bahubalindruni2013transparent}
Bahubalindruni,~P.~G.; Barquinha,~P.; Duarte,~C.; de~Oliveira,~P.~G.; Martins,~R.; Fortunato,~E.; others \href{https://ieeexplore.ieee.org/document/6571236}{Transparent current mirrors with a-GIZO TFTs: Neural modeling, simulation and fabrication}. \emph{J. Disp. Technol.} \textbf{2013}, \emph{9}, 1001--1006\relax
\mciteBstWouldAddEndPuncttrue
\mciteSetBstMidEndSepPunct{\mcitedefaultmidpunct}
{\mcitedefaultendpunct}{\mcitedefaultseppunct}\relax
\EndOfBibitem
\bibitem[Luo \latin{et~al.}(2023)Luo, Abidian, Ahn, Akinwande, Andrews, Antonietti, Bao, Berggren, Berkey, Bettinger, Chen, Chen, Cheng, Cheng, Choi, Chortos, Dagdeviren, Dauskardt, Di, Dickey, Duan, and \textit{et al.}]{Luo_2023}
Luo,~Y. \latin{et~al.}  \href{https://doi.org/10.1021/acsnano.2c12606}{Technology Roadmap for Flexible Sensors}. \emph{ACS Nano} \textbf{2023}, \emph{17}, 5211--5295\relax
\mciteBstWouldAddEndPuncttrue
\mciteSetBstMidEndSepPunct{\mcitedefaultmidpunct}
{\mcitedefaultendpunct}{\mcitedefaultseppunct}\relax
\EndOfBibitem
\bibitem[Gruber \latin{et~al.}(1997)Gruber, Dr{\"a}benstedt, Tietz, Fleury, Wrachtrup, and von Borczyskowski]{Gruber:1997}
Gruber,~A.; Dr{\"a}benstedt,~A.; Tietz,~C.; Fleury,~L.; Wrachtrup,~J.; von Borczyskowski,~C. \href{https://doi.org/10.1126/science.276.5321.2012}{Scanning Confocal Optical Microscopy and Magnetic Resonance on Single Defect Centers}. \emph{Science} \textbf{1997}, \emph{276}, 2012--2014\relax
\mciteBstWouldAddEndPuncttrue
\mciteSetBstMidEndSepPunct{\mcitedefaultmidpunct}
{\mcitedefaultendpunct}{\mcitedefaultseppunct}\relax
\EndOfBibitem
\bibitem[Doherty \latin{et~al.}(2013)Doherty, Manson, Delaney, Jelezko, Wrachtrup, and Hollenberg]{Doherty:2013}
Doherty,~M.~W.; Manson,~N.~B.; Delaney,~P.; Jelezko,~F.; Wrachtrup,~J.; Hollenberg,~L.~C. \href{https://doi.org/10.1016/j.physrep.2013.02.001}{The nitrogen-vacancy colour centre in diamond}. \emph{Physics Reports} \textbf{2013}, \emph{528}, 1--45\relax
\mciteBstWouldAddEndPuncttrue
\mciteSetBstMidEndSepPunct{\mcitedefaultmidpunct}
{\mcitedefaultendpunct}{\mcitedefaultseppunct}\relax
\EndOfBibitem
\bibitem[Roth \latin{et~al.}(1989)Roth, Sepulveda, and Wikswo~Jr]{roth}
Roth,~B.~J.; Sepulveda,~N.~G.; Wikswo~Jr,~J.~P. \href{http://dx.doi.org/10.1063/1.342549}{Using a magnetometer to image a two-dimensional current distribution}. \emph{Jour. of App. Physics} \textbf{1989}, \emph{65}, 361--372\relax
\mciteBstWouldAddEndPuncttrue
\mciteSetBstMidEndSepPunct{\mcitedefaultmidpunct}
{\mcitedefaultendpunct}{\mcitedefaultseppunct}\relax
\EndOfBibitem
\bibitem[Tetienne \latin{et~al.}(2019)Tetienne, Dontschuk, Broadway, Lillie, Teraji, Simpson, Stacey, and Hollenberg]{Tetienne:2019}
Tetienne,~J.-P.; Dontschuk,~N.; Broadway,~D.~A.; Lillie,~S.~E.; Teraji,~T.; Simpson,~D.~A.; Stacey,~A.; Hollenberg,~L. C.~L. \href{https://link.aps.org/doi/10.1103/PhysRevB.99.014436}{Apparent delocalization of the current density in metallic wires observed with diamond nitrogen-vacancy magnetometry}. \emph{Phys. Rev. B} \textbf{2019}, \emph{99}, 014436\relax
\mciteBstWouldAddEndPuncttrue
\mciteSetBstMidEndSepPunct{\mcitedefaultmidpunct}
{\mcitedefaultendpunct}{\mcitedefaultseppunct}\relax
\EndOfBibitem
\bibitem[Broadway \latin{et~al.}(2020)Broadway, Lillie, Scholten, Rohner, Dontschuk, Maletinsky, Tetienne, and Hollenberg]{Broadway:2020}
Broadway,~D.; Lillie,~S.; Scholten,~S.; Rohner,~D.; Dontschuk,~N.; Maletinsky,~P.; Tetienne,~J.-P.; Hollenberg,~L. \href{https://link.aps.org/doi/10.1103/PhysRevApplied.14.024076}{Improved Current Density and Magnetization Reconstruction Through Vector Magnetic Field Measurements}. \emph{Phys. Rev. Appl.} \textbf{2020}, \emph{14}, 024076\relax
\mciteBstWouldAddEndPuncttrue
\mciteSetBstMidEndSepPunct{\mcitedefaultmidpunct}
{\mcitedefaultendpunct}{\mcitedefaultseppunct}\relax
\EndOfBibitem
\bibitem[Abrahams \latin{et~al.}(2021)Abrahams, Scholten, Healey, Robertson, Dontschuk, Lim, Johnson, Simpson, Hollenberg, and Tetienne]{Abrahams_2021}
Abrahams,~G.~J.; Scholten,~S.~C.; Healey,~A.~J.; Robertson,~I.~O.; Dontschuk,~N.; Lim,~S.~Q.; Johnson,~B.~C.; Simpson,~D.~A.; Hollenberg,~L. C.~L.; Tetienne,~J.-P. \href{https://doi.org/10.1063/5.0073320}{An integrated widefield probe for practical diamond nitrogen-vacancy microscopy}. \emph{Appl. Phys. Lett.} \textbf{2021}, \emph{119}, 254002\relax
\mciteBstWouldAddEndPuncttrue
\mciteSetBstMidEndSepPunct{\mcitedefaultmidpunct}
{\mcitedefaultendpunct}{\mcitedefaultseppunct}\relax
\EndOfBibitem
\end{mcitethebibliography}
\end{document}